\documentclass[mathpazo]{cicp}
\usepackage{graphicx}
\usepackage{subfigure}
\usepackage{bm}
\usepackage{epsfig}
\usepackage{color}
\begin{document}
%%%%% title : short title may not be used but TITLE is required.
% \title{TITLE}
% \title[short title]{TITLE}
\title{An efficient high-order gas-kinetic scheme with hybrid WENO-AO method for the Euler and Navier-Stokes solutions}

%%%%% author(s) :
% single author:
% \author[name in running head]{AUTHOR\corrauth}
% [name in running head] is NOT OPTIONAL, it is a MUST.
% Use \corrauth to indicate the corresponding author.
% Use \email to provide email address of author.
% \footnote and \thanks are not used in the heading section.
% Another acknowlegments/support of grants, state in Acknowledgments section
% \section*{Acknowledgments}
%\author[O.~Author]{Junlei Mu\corrauth}
%\address{School of Mathematical Sciences, Beijing Normal University,
%Beijing 100875, P.R. China}
%\email{{\tt author@email} (O.~Author)}

\author[Mu J L et.~al.]{Junlei Mu\affil{1},
     Congshan Zhuo\affil{1,2}\comma\corrauth, Qingdian Zhang\affil{1}, Sha Liu\affil{1,2}, and Chengwen Zhong\affil{1,2}}
\address{\affilnum{1}\ School of Aeronautics,
			 Northwestern Polytechnical University, Xi'an, Shaanxi 710072, China \\
          \affilnum{2}\ National Key Laboratory of Science and Technology on Aerodynamic Design and Research, Northwestern Polytechnical University, Xi'an, Shaanxi 710072, China}
 \emails{{\tt 2021200064@mail.nwpu.edu.cn} (J.~Mu), 
 	{\tt zhuocs@nwpu.edu.cn} (C.~Zhuo), 
 	{\tt zhangqd@mail.nwpu.edu.cn} (Q.~Zhang),
 	{\tt shaliu@nwpu.edu.cn} (S.~Liu), {\tt zhongcw@nwpu.edu.cn} (C.~Zhong)}
% multiple authors:
% Note the use of \affil and \affilnum to link names and addresses.
% The author for correspondence is marked by \corrauth.
% use \emails to provide email addresses of authors
% e.g. below example has 3 authors, first author is also the corresponding
%      author, author 1 and 3 having the same address.
% \author[Zhang Z R et.~al.]{Zhengru Zhang\affil{1}\comma\corrauth,
%       Author Chan\affil{2}, and Author Zhao\affil{1}}
% \address{\affilnum{1}\ School of Mathematical Sciences,
%          Beijing Normal University,
%          Beijing 100875, P.R. China. \\
%           \affilnum{2}\ Department of Mathematics,
%           Hong Kong Baptist University, Hong Kong SAR}
% \emails{{\tt zhang@email} (Z.~Zhang), {\tt chan@email} (A.~Chan),
%          {\tt zhao@email} (A.~Zhao)}
% \footnote and \thanks are not used in the heading section.
% Another acknowlegments/support of grants, state in Acknowledgments section
% \section*{Acknowledgments}

%%%%% Begin Abstract %%%%%%%%%%%
\begin{abstract}
The high-order gas-kinetic scheme (HGKS) features good robustness, high efficiency and satisfactory accuracy,the performaence of which can be further improved combined with  WENO-AO (WENO with adaptive order) scheme for reconstruction. To reduce computational costs in the reconstruction procedure, this paper proposes to combine HGKS with a hybrid WENO-AO scheme. The hybrid WENO-AO scheme reconstructs target variables using upwind linear approximation directly if all extreme points of the reconstruction polynomials for these variables are outside the large stencil. Otherwise, the WENO-AO scheme is used. Unlike combining the hybrid WENO scheme with traditional Riemann solvers, the troubled cell indicator of the hybrid WENO-AO method is fully utilized in the spatial reconstruction process of HGKS. During normal and tangential reconstruction, the gas-kinetic scheme flux not only needs to reconstruct the conservative variables on the left and right interfaces but also to reconstruct the derivative terms of the conservative variables. By reducing the number of times that the WENO-AO scheme is used, the calculation cost is reduced. The high-order gas-kinetic scheme with the hybrid WENO-AO method retains original robustness and accuracy of the WENO5-AO GKS, while exhibits higher computational efficiency.
\end{abstract}
%%%%% end %%%%%%%%%%%

%%%%% AMS/PACs/Keywords %%%%%%%%%%%
%\pac{}
\ams{52B10, 65D18, 68U05, 68U07}
\keywords{Gas-kinetic scheme, Hybrid WENO-AO scheme, High-order finite volume scheme, Extreme point.}

%%%% maketitle %%%%%
\maketitle

%%%% Start %%%%%%
\section{Introduction}
\label{sec1}
	The development of aerospace technology has accentuated the need to describe complex and detailed flow fields around high-speed aircrafts efficiently. The research focus in computational fluid dynamics (CFD) ~\cite{HCM} has shifted towards high-resolution and high-order schemes. In recent years, emphasis has been placed on high-order finite volume methods~\cite{AGBS}, discontinuous Galerkin (DG) methods, and correction procedures via reconstruction (CPR) for numerical simulations involving complex geometric meshes. The gas-kinetic scheme is a numerical method that provides solutions to the Euler and Navier-Stokes equations within the finite volume framework. The high-order gas-kinetic scheme (HGKS) has demonstrated excellent performance in compressible flows~\cite{AHAG,art6}. It has also been successfully extended into DG~\cite{AMHD} methods and CPR~\cite{ATGC} frameworks.\par
	The gas-kinetic scheme employs a time-evolving gas distribution function to reveal the multi-scale physical flow mechanism, encompassing kinetic particle transport to hydrodynamic wave propagation~\cite{PEFH}. A one-stage third-order high-order gas-kinetic scheme was applied for subsonic to supersonic flow tests under the three-dimensional hybrid unstructured grid~\cite{ATGS2}. To solve the Euler equations, the efficient high-order gas-kinetic scheme (EHGKS) uses both HGKS and Lax-Wendroff method and can be generalized to arbitrary accuracy~\cite{AEHG1}. A two-stage fourth-order HGKS scheme was proposed by Pan et al.~\cite{AEAA} based on the two-stage fourth-order generalized Riemann problem (GRP) scheme~\cite{ATFO1}, becoming increasingly popular. The combination of multi-stage multi-derivative method (MSMD) and HGKS using second-order and third-order gas distribution functions resulted in two-stage fifth-order and three-stage fifth-order HGKS~\cite{AFOH}, with higher efficiency than the traditional Riemann solver~\cite{RSNU} with Runge-Kutta (RK) time stepping method. Fifth-order WENO scheme was proposed by Jiang and Shu~\cite{EIOW}, providing a general framework for designing smoothness indicators and nonlinear weights. For spatial reconstruction, the classic HGKS with WENO reconstruction~\cite{PEFH} is similar to the traditional high-order finite volume method (FVM)~\cite{FWSF}. But the classic HGKS with WENO reconstruction~\cite{PEFH} better captures shear instability due to the multidimensional properties of the gas distribution function~\cite{AMGB}. Using fifth-order WENO-JS or WENO-Z reconstruction, non-equilibrium state and equilibrium state in the gas distribution function were generally reconstructed separately. The WENO adaptive order (WENO-AO) method proposed by Balsara~\cite{AECO} recently was for HGKS spatial reconstruction, overcoming shortcomings of the classic HGKS~\cite{PEFH}. Compared to the classic WENO5-GKS scheme, HGKS with WENO-AO reconstruction has shown significant improvements in performance and no false overshoots or undershoots in certain test cases~\cite{PEFH}. In addition, the high-order gas-kinetic scheme and high-order gas-kinetic compact schemes~\cite{CHGS,ACHG2,AHRB} are preferred on unstructured meshes due to their higher resolution and stability.\par
	A fundamental concept of WENO schemes is to use a linear combination of lower order numerical fluxes or reconstructions to obtain a higher order approximation. For systems, WENO schemes employ a local characteristic decomposition method to avoid spurious oscillations. However, computing the nonlinear weights and local characteristic decompositions is computationally expensive. Pirozzoli~\cite{CHCS2} developed an efficient hybrid scheme that combines the WENO scheme for discontinuous regions with a compact upwind scheme for smooth regions. Hill and Pullin~\cite{HTCD} proposed a scheme that integrates the tuned center difference method with the WENO scheme to automatically implement nonlinear weights in smooth regions away from shocks. Li and Qiu~\cite{ACOT} designed a hybrid WENO scheme using switching principles and Runge-Kutta time discretization. Huang and Qiu~\cite{HWSW2} proposed a hybrid WENO scheme using the Lax-Wendroff time discretization program. However, different troubled cell indicators may have varying effects on hybrid WENO schemes, and these indicators need to adjust parameters based on different problems to better maintain non-oscillation near discontinuities and reduce computational costs. Zhu and Qiu~\cite{ANHW} addressed this issue by proposing a new simple switching principle that uses different reconstruction methods by identifying the positions of all extreme points of a large reconstructed polynomial with numerical fluxes. Chen and Huang~\cite{ODHW} improved upon the troubled cell indicator, eliminating the difficulty of finding extreme points of fourth-degree polynomials that can be extended to higher-order hybrid WENO schemes. This study focuses on the hybrid WENO-AO method proposed by Chen and Huang.\par
	This paper applies the hybrid WENO-AO scheme presented in~\cite{ODHW} to further reduce computation costs and increase calculation speed while maintaining the robustness and high accuracy of the existing HGKS with WENO-AO method. The benefits of the hybrid WENO-AO scheme for HGKS are as follows. First, the HGKS with the hybrid WENO-AO method retains all the advantages of the HGKS with WENO-AO method~\cite{PEFH}. Second, computing characteristic decompositions and nonlinear weights in the WENO scheme is expensive. The hybrid WENO-AO scheme employs upwind linear reconstruction in smooth regions, which avoids the need for characteristic decompositions and calculations of nonlinear weights. Third, the WENO-AO scheme requires high-order smoothness indicators of large stencils to obtain nonlinear weights. Therefore, it is meaningful to replace WENO-AO reconstruction with linear upwind reconstruction in smooth regions to reduce the calculation cost of smoothness indicators. Additionally, for the WENO5-AO GKS method, calculating nonlinear weights is computationally expensive due to the multiple uses of the WENO-AO scheme during normal and tangential reconstruction for interface flux. For the HGKS with the hybrid WENO method, the new scheme utilizes upwind linear approximation in smooth regions, which reduces the computation of smoothness indicators, nonlinear weights, and local characteristic decompositions, saving CPU time. In brief, the HGKS with hybrid WENO scheme maintains the robustness of the HGKS with WENO-AO scheme~\cite{PEFH} and demonstrates high efficiency and practicality.\par
	This paper is organized as follows. The second section presents the classic two-stage fourth-order temporal discretization and GKS flux solver. Then, the HGKS with hybrid WENO-AO scheme is introduced, along with a reconstruction procedure for obtaining the non-equilibrium and equilibrium states of the gas distribution function in 1D and 2D. In the third section, numerical tests are conducted to evaluate the accuracy, efficiency, robustness, and shock capture capabilities of the HGKS with hybrid WENO method. Finally, the last section provides a summary of the high-order gas-kinetic scheme with hybrid WENO-AO method.

\section{High-order gas-kinetic scheme with hybrid WENO-AO}
\label{sec2}
In this section, we first provide a brief introduction to the gas-kinetic flux solver and the two-stage fourth-order temporal discretization~\cite{AEAA}. Next, we describe the HGKS with hybrid WENO-AO spatial reconstruction in detail. Additionally, we conduct a simple analysis of the accuracy and computational costs of the hybrid WENO-AO reconstruction compared to the original WENO-AO reconstruction within the framework of HGKS.\par
  As the finite volume scheme, the conservation laws can be written as
$${{\mathbf{W}}_{t}}=-\nabla \cdot F\left( \mathbf{W} \right),$$
$$\mathbf{W}\left( 0,x \right)={{\mathbf{W}}_{0}}\left( x \right),x\in \Omega \subseteq \mathbb{R},$$
where $\mathbf{W}$ is the conservative variables and $F$ is the corresponding flux. With the spatial discretization ${{\mathbf{W}}^{h}}$ and appropriate evaluation  $-\nabla \cdot F\left( \mathbf{W} \right)$. The original partial differential equation (PDE) became the ordinary differential equation (ODE).
\begin{equation}\label{equ1}
	\mathbf{W}_{t}^{h}=\mathcal{L}\left( {{\mathbf{W}}^{h}} \right),t={{t}_{n}},
\end{equation}
where $\mathcal{L}\left( {{\mathbf{W}}^{h}} \right)$ is the spatial operator of flux.
\subsection{Gas-kinetic flux solver}
\label{sec2.1}
The two-dimensional Boltzmann-BGK equation is written as
\begin{equation}
	{{f}_{t}}+u{{f}_{x}}+v{{f}_{y}}=\frac{g-f}{\tau },
\end{equation}
where $\mathbf{u}=\left( u,v \right)$ is the particle velocity, $\tau $ is the collision time. $f$ is the gas distribution function and $g$ is the equilibrium state~\cite{AMFC}.\par
  The equilibrium state $g$ is a Maxwellian distribution,
\begin{equation}
	g=\rho {{\left( \frac{\lambda }{\pi } \right)}^{\frac{K+2}{2}}}{{e}^{-\lambda \left( {{\left( u-U \right)}			^{2}}+{{\left( v-V \right)}^{2}}+{{\xi }^{2}} \right)}},
\end{equation}
where $\rho $ is the density and $\lambda =m/2kT$, where $m$ is the molecular mass, $k$ is the Boltzmann constant, and $T$ is the temperature. $U$ and $V$ are the macroscopic velocities in the x-direction and y-direction. For a 2D flow, the internal variable $\xi $  include the random particle motion in the z-direction, and the total number of degrees of freedom $K=\left( 5-3\gamma  \right)/\left( \gamma -1 \right)+1$. $\gamma $ is the specific heat ratio. In the equilibrium state, the internal variable ${{\xi }^{2}}=\xi _{1}^{2}+\xi _{2}^{2}+\cdots +\xi _{K}^{2}$. The relation between the conservative variables $\left( \rho ,\rho U,\rho V,\rho E \right)$ and the distribution function $f$ is
\begin{equation}
\left( \begin{aligned}
  & \rho  \\ 
 & \rho U \\ 
 & \rho V \\ 
 & \rho E \\ 
\end{aligned} 
\right)=\int{{{\psi }_{\alpha }}fd\Xi ,\alpha =1,2,3,4},
\end{equation}
where ${{\psi }_{\alpha }} ={{\left( {{\psi }_{1}},{{\psi }_{2}},{{\psi }_{3}},{{\psi }_{4}} \right)}^{T}}={{\left( 1,u,v,\frac{1}{2}\left( {{u}^{2}}+{{v}^{2}}+{{\xi }^{2}} \right) \right)}^{T}}$, $d\Xi =dudvd{{\xi }_{1}}\ldots d{{\xi }_{K}}$. The collision term satisfies the following compatibility condition
\begin{equation}
	\int{\frac{g-f}{\tau }}\psi d\Xi =0.
\end{equation}
Based on the Chapman-Enskog expansion for Boltzmann-BGK equation~\cite{KFVS1,ANDV}, the gas distribution function in the continuum regime can be expanded as
$$f=g-\tau {{D}_{u}}g+\tau {{D}_{u}}\left( \tau {{D}_{u}} \right)g-\tau {{D}_{u}}\left[ \tau {{D}_{u}}\left( \tau {{D}_{u}} \right)g \right]+\cdots, $$
where ${{D}_{u}}=\partial /\partial t+u\cdot \nabla $. By truncating on different orders of $\tau $, the corresponding macroscopic equations can be derived~\cite{AMDO}. For the Euler equations, the zeroth order truncation is taken, i.e. $f=g$. For the Navier-Stokes equations, when the first order truncation is used, the distribution function is
$$f=g-\tau \left( u{{g}_{x}}+v{{g}_{y}}+{{g}_{t}} \right).$$
The solution $f$ of the Boltzmann-BGK equation at a cell interface ${{x}_{j+1/2}}$ is
\begin{equation}\label{equ6}
f\left( {{x}_{i+1/2}},t,u,v,\xi  \right)=\frac{1}{\tau }\int_{0}^{t}{g\left( {x}',{t}',u,v,\xi  \right)}{{e}^{-\left( t-{t}' \right)/\tau }}d{t}'+{{e}^{-t/\tau }}{{f}_{0}}\left( -uty-vt,u,v,\xi  \right),
\end{equation}
where ${x}'={{x}_{i+1/2}}-u\left( t-{t}' \right)$ and ${y}'=y-v\left( t-{t}' \right)$ are the particle trajectory, ${{x}_{i+1/2}}=0$ is the location of the cell interface. ${{f}_{0}}$ is the initial gas distribution function $f$ at the beginning of each time step $\left( t=0 \right)$. The integral solution simulates the physical process from the free transport of particles in the kinetic scale physics ${{f}_{0}}$ to the evolution of hydrodynamic flow in the integral of $g$ term. The flow behavior at the cell interface depends on the ratio of the time step to the local particle collision time $\Delta t/\tau $.\par
The initial gas distribution function ${{f}_{0}}$ with considering the first order non-equilibrium part is expressed as follows:
\begin{equation}
	{{f}_{0}}=\left\{ \begin{aligned}
  & {{g}^{l}}\left( 1+a_{1}^{l}x+a_{2}^{l}y-\tau \left( a_{1}^{l}u+a_{2}^{l}v+{{A}^{l}} \right) \right),x\le 0, \\ 
 & {{g}^{r}}\left( 1+a_{1}^{r}x+a_{2}^{r}y-\tau \left( a_{1}^{r}u+a_{2}^{r}v+{{A}^{r}} \right) \right),x>0. \\ 
\end{aligned} \right.
\end{equation}
The relation between the terms of the ${{f}_{0}}$ and macroscopic variables ${Q}$ is
\begin{equation}
	\left\langle a_{1}^{k} \right\rangle =\frac{\partial {{Q}_{k}}}{\partial {{\mathbf{n}}_{x}}},\left\langle a_{2}^{k} \right\rangle =\frac{\partial {{Q}_{k}}}{\partial {{\mathbf{n}}_{y}}},
\end{equation}
where $k=l,r$ and $\left\langle \ldots  \right\rangle $ are the moments of the equilibrium $g$ and defined by
$$\left\langle \ldots  \right\rangle =\int{g\left( \ldots  \right)}\psi d\Xi. $$
The coefficients can be determined by the spatial derivatives of macroscopic flow variables and the compatibility condition as follows
\begin{equation}
	\begin{aligned}
  & \int{\psi {{g}_{0}}d\Xi ={{Q}_{0}}=}\int_{u>0}{\psi {{g}_{l}}d\Xi +}\int_{u<0}{\psi {{g}_{r}}d\Xi }, \\ 
 & \left\langle {{{\bar{a}}}_{1}} \right\rangle =\frac{\partial {{Q}_{0}}}{\partial {{\mathbf{n}}_{x}}}=\int_{u>0}{\psi a_{1}^{l}{{g}_{l}}d\Xi +}\int_{u<0}{\psi a_{1}^{r}{{g}_{r}}d\Xi }, \\ 
 & \left\langle {{{\bar{a}}}_{2}} \right\rangle =\frac{\partial {{Q}_{0}}}{\partial {{\mathbf{n}}_{y}}}=\int_{u>0}{\psi a_{2}^{l}{{g}_{l}}d\Xi +}\int_{u<0}{\psi a_{2}^{r}{{g}_{r}}d\Xi }, \\ 
 & \left\langle a_{1}^{k}u+a_{2}^{k}v+{{A}^{k}} \right\rangle =0,\left\langle {{{\bar{a}}}_{1}}u+{{{\bar{a}}}_{2}}v+\bar{A} \right\rangle =0, \\ 
\end{aligned}\
\end{equation}
where $a\equiv \left( \partial g/\partial x \right)/g={{g}_{x}}/g,A\equiv \left( \partial g/\partial t \right)/g={{g}_{t}}/g$, After obtaining all the required terms, substitute these terms into Eq. (\ref{equ6}) and solve it can obtain the gas distribution function on the interface~\cite{AGBS}.
\begin{equation}\label{equ10}
	\begin{aligned}
  & f\left( {{x}_{i+1/2}},t,u,v,\xi  \right)=\left( 1-{{e}^{-t/\tau }} \right){{g}_{0}}+\left( \left( t+\tau  \right){{e}^{-t/\tau }}-\tau  \right)\left( {{{\bar{a}}}_{1}}u+{{{\bar{a}}}_{2}}v \right){{g}_{0}} \\ 
 & +\left( t-\tau +\tau {{e}^{-t/\tau }} \right)\bar{A}{{g}_{0}} \\ 
 & +{{e}^{-t/\tau }}{{g}_{r}}\left[ 1-\left( \tau +t \right)\left( a_{1}^{r}u+a_{2}^{r}v \right)-\tau {{A}^{r}} \right]H\left( u \right) \\ 
 & +{{e}^{-t/\tau }}{{g}_{l}}\left[ 1-\left( \tau +t \right)\left( a_{1}^{l}u+a_{2}^{l}v \right)-\tau {{A}^{l}} \right]\left( 1-H\left( u \right) \right),  
\end{aligned}
\end{equation}
where $H(u)$ is the Heaviside function. Derivations related to gas kinetic scheme can be found in ~\cite{AGBS}.\par
 Finally, the gas-kinetic numerical fluxes in the x-direction on the cell interface can be computed as
\begin{equation}
	{{F}_{i+1/2}}\left( {{\mathbf{W}}^{n}},t \right)=\int{u\left( \begin{aligned}
  & 1 \\ 
 & u \\ 
 & v \\ 
 & \frac{1}{2}\left( {{u}^{2}}+{{v}^{2}}+{{\xi }^{2}} \right) \\ 
\end{aligned} \right)f\left( {{x}_{j+1/2}},t,u,v,\xi  \right)d\Xi .}
\end{equation}
By integrating the above equation in the time interval $\left[ {{t}_{n}},\ {{t}_{n}}+\Delta t \right]$, the total mass, momentum, and energy transport can be written as
\begin{equation}\label{equ12}
{{\mathbb{F}}_{i+1/2}}\left( {{\mathbf{W}}^{n}},\delta  \right)=\int_{{{t}_{n}}}^{{{t}_{n}}+\delta }{{{F}_{i+1/2}}\left( {{\mathbf{W}}^{n}},t \right)}dt
\end{equation}
More details can be found in ~\cite{AGBS}.

\subsection{Two-stage fourth-order temporal discretization}
\label{sec2.2}
There are many ways to solve ODE Eq. (\ref{equ1}). Here we define
$$\mathbf{W}_{t}^{^{\left( m \right)}}\left( {{t}^{n}} \right)=\frac{{{d}^{m}}{{\mathbf{W}}^{n}}}{d{{t}^{m}}}=\frac{{{d}^{m-1}}\mathcal{L}\left( {{\mathbf{W}}^{n}} \right)}{d{{t}^{m-1}}}={{\mathcal{L}}^{^{m-1}}}.$$
The two-stage fourth-order time marching scheme is used to solve the initial value problem Eq. (\ref{equ1}) and is written as
\begin{equation}
\begin{aligned}
  & {{\mathbf{W}}^{*}}={{\mathbf{W}}^{n}}+\frac{1}{2}\Delta t\mathcal{L}\left( {{\mathbf{W}}^{n}} \right)+\frac{1}{8}\Delta {{t}^{2}}\frac{\partial }{\partial t}\mathcal{L}\left( {{\mathbf{W}}^{n}} \right) \\ 
 & {{\mathbf{W}}^{n+1}}={{\mathbf{W}}^{n}}+\Delta t\mathcal{L}\left( {{\mathbf{W}}^{n}} \right)+\frac{1}{6}\Delta {{t}^{2}}\left( \frac{\partial }{\partial t}\mathcal{L}\left( {{\mathbf{W}}^{n}} \right)+2\frac{\partial }{\partial t}\mathcal{L}\left( {{\mathbf{W}}^{*}} \right) \right), \\ 
\end{aligned}
\end{equation}
where $\partial \mathcal{L}\left( \mathbf{W} \right)/\partial t$ is the time derivative of the spatial operator. If the time derivatives of ${{\mathcal{L}}^{\left( n \right)}}$ up to $\left( n-1 \right)$ can be given, a $n$th-order time marching scheme can be constructed directly. Usually only a few low-order derivative is used to solve nonlinear system problems, such as $\mathcal{L}$ for approximating the Riemann solver. For hyperbolic conservation law, the two-stage fourth-order temporal discretization is derived in ~\cite{ATFO1} and ${{\mathcal{L}}^{\left( 1 \right)}}$ for the generalized Riemann problem (GRP) solver. In the establishment of high-order gas-kinetic schemes, ${{\mathcal{L}}^{\left( 1 \right)}}$ and ${{\mathcal{L}}^{\left( 2 \right)}}$ are often used by HGKS for 2nd-order GKS flux functions and 3rd-order GKS flux functions respectively~\cite{PMAN,ACFG,ATFG2}. However, higher-order derivatives are prohibited due to the complexity of the HGKS flux function. Another multi-stage multi-derivative method (MSMD) was proposed with HGKS and continues to be used. Similar to the Runge-Kutta (RK) time marching scheme, the MSMD introduce middle stages. The update at ${{t}^{n+1}}$ becomes a linear combination of $\mathcal{L}$ and their derivatives in multiple stages. Because of the use of $\mathcal{L}$ and their derivatives, MSMD can achieve the required time accuracy with fewer middle stages than RK method. This paper mainly focuses on the two-stage fourth-order temporal discretization scheme (S2O4), and more HGKS with MSMD method is detailed and analyzed in ~\cite{AFOH}.\par
Eq. (\ref{equ10}) provides a time-dependent gas distribution function. And the flux for the macroscopic flow variables can be calculated by Eq. (\ref{equ12}). The flux of the two-stage fourth-order GKS in the time interval $\left[ {{t}_{n}},\ {{t}_{n}}+\Delta t \right]$ needs the ${{F}_{i\pm 1/2}}\left( \mathbf{W} \right)$ and ${{\partial }_{t}}{{F}_{i\pm 1/2}}\left( \mathbf{W} \right)$ at both ${{t}_{n}}$ and ${{t}_{*}}={{t}_{n}}+\Delta t/2$. The followings mainly introduce the two-stage algorithm.\\
(1)	With the reconstruction at ${{t}_{n}}$, the x-direction flux ${{\mathbb{F}}_{i+1/2,j}}\left( {{\mathbf{W}}^{n}},\Delta t \right)$, ${{\mathbb{F}}_{i+1/2,j}}\left( {{\mathbf{W}}^{n}},\Delta t/2 \right)$ and y-direction flux ${{\mathbb{G}}_{i,j+1/2}}\left( {{\mathbf{W}}^{n}},\Delta t \right)$ and ${{\mathbb{G}}_{i,j+1/2}}\left( {{\mathbf{W}}^{n}},\Delta t/2 \right)$ in the time interval $\left[ {{t}_{n}},\ {{t}_{n}}+\frac{\Delta t}{2} \right]$ can be evaluated by Eq. (\ref{equ12}).\\
(2)	In the time interval $\left[ {{t}_{n}},\ {{t}_{n}}+\Delta t \right]$, the flux is expanded as the linear form as
	\begin{equation}
{{F}_{i+1/2,j}}\left( {{\mathbf{W}}^{n}},{{t}_{n}} \right)=F_{_{i+1/2,j}}^{n}+{{\partial }_{t}}F_{_{i+1/2,j}}^{n}\left( t-{{t}_{n}} \right).
\end{equation}
The terms $F_{_{i+1/2,j}}^{{}}\left( {{\mathbf{W}}^{n}},{{t}_{n}} \right)$ and $\partial tF_{_{i+1/2,j}}^{{}}\left( {{\mathbf{W}}^{n}},{{t}_{n}} \right)$ can be determined as
	\begin{equation}	 
{{F}_{i+1/2,j}}\left( {{\mathbf{W}}^{n}},{{t}_{n}} \right)\Delta t+\frac{1}{2}{{\partial }_{t}}{{F}_{i+1/2,j}}\left( {{\mathbf{W}}^{n}},{{t}_{n}} \right)\Delta {{t}^{2}}={{\mathbb{F}}_{i+1/2,j}}\left( {{\mathbf{W}}^{n}},\Delta t \right),
\end{equation}
	\begin{equation}	 
\frac{1}{2}{{F}_{i+1/2,j}}\left( {{\mathbf{W}}^{n}},{{t}_{n}} \right)\Delta t+\frac{1}{8}{{\partial }_{t}}{{F}_{i+1/2,j}}\left( {{\mathbf{W}}^{n}},{{t}_{n}} \right)\Delta {{t}^{2}}={{\mathbb{F}}_{i+1/2,j}}\left( {{\mathbf{W}}^{n}},\Delta t/2 \right).
\end{equation}
The terms $F_{_{i+1/2,j}}^{{}}\left( {{\mathbf{W}}^{n}},{{t}_{n}} \right)$ and $\partial tF_{_{i+1/2,j}}^{{}}\left( {{\mathbf{W}}^{n}},{{t}_{n}} \right)$ can be computed by solving this linear system as
\begin{equation}
{{F}_{i+1/2,j}}\left( {{\mathbf{W}}^{n}},{{t}_{n}} \right)=\left( 4{{\mathbb{F}}_{i+1/2,j}}\left( {{\mathbf{W}}^{n}},\Delta t/2 \right)-{{\mathbb{F}}_{i+1/2,j}}\left( {{\mathbf{W}}^{n}},\Delta t \right) \right)/\Delta t,
\end{equation}
\begin{equation}
{{\partial }_{t}}{{F}_{i+1/2,j}}\left( {{\mathbf{W}}^{n}},{{t}_{n}} \right)=4\left( {{\mathbb{F}}_{i+1/2,j}}\left( {{\mathbf{W}}^{n}},\Delta t \right)-2{{\mathbb{F}}_{i+1/2,j}}\left( {{\mathbf{W}}^{n}},\Delta t/2 \right) \right)/\Delta {{t}^{2}}.
\end{equation}
Similar to $F_{_{i+1/2,j}}^{{}}\left( {{\mathbf{W}}^{n}},{{t}_{n}} \right)$ and $\partial tF_{_{i+1/2,j}}^{{}}\left( {{\mathbf{W}}^{n}},{{t}_{n}} \right)$, the $G_{_{i,j+1/2}}^{{}}\left( {{\mathbf{W}}^{n}},{{t}_{n}} \right)$ and $\partial tG_{_{i,j+1/2}}^{{}}\left( {{\mathbf{W}}^{n}},{{t}_{n}} \right)$ can be computed from ${{\mathbb{G}}_{i,j+1/2}}\left( {{\mathbf{W}}^{n}},\Delta t \right)$ and ${{\mathbb{G}}_{i,j+1/2}}\left( {{\mathbf{W}}^{n}},\Delta t/2 \right)$.\\
(3)	Update $\mathbf{W}_{ij}^{*}$ at ${{t}_{*}}={{t}_{n}}+\Delta t/2$ by
	 \begin{equation}
\begin{aligned}
  & \mathbf{W}_{ij}^{*}=\mathbf{W}_{ij}^{n}-\frac{1}{\Delta x}\left[ {{\mathbb{F}}_{i+1/2,j}}\left( {{\mathbf{W}}^{n}},\Delta t/2 \right)-{{\mathbb{F}}_{i-1/2,j}}\left( {{\mathbf{W}}^{n}},\Delta t/2 \right) \right] \\ 
 & -\frac{1}{\Delta y}\left[ {{\mathbb{G}}_{i,j+1/2}}\left( {{\mathbf{W}}^{n}},\Delta t/2 \right)-{{\mathbb{G}}_{i,j-1/2}}\left( {{\mathbf{W}}^{n}},\Delta t/2 \right) \right].  
\end{aligned}
\end{equation}

Again through the step (1) and step (2) to compute the middle stage value, we can get $\partial tF_{_{i+1/2,j}}^{{}}\left( {{\mathbf{W}}^{*}},{{t}_{*}} \right)$ and $\partial tG_{_{i,j+1/2}}^{{}}\left( {{\mathbf{W}}^{*}},{{t}_{*}} \right)$ in the time interval $\left[ {{t}_{*}},\ {{t}_{*}}+\Delta t \right]$.\\
(4)	The numerical fluxes $\mathcal{F}_{i+1/2,j}^{n}$ and $\mathcal{G}_{i+1/2,j}^{n}$ can be computed by 
	 \begin{equation}
\mathcal{F}_{i+1/2,j}^{n}={{F}_{i+1/2,j}}\left( {{\mathbf{W}}^{n}},{{t}_{n}} \right)+\frac{\Delta t}{6}\left[ {{\partial }_{t}}{{F}_{i+1/2,j}}\left( {{\mathbf{W}}^{n}},{{t}_{n}} \right)+2{{\partial }_{t}}{{F}_{i+1/2,j}}\left( {{\mathbf{W}}^{*}},{{t}_{*}} \right) \right],
\end{equation}
	 \begin{equation}
\mathcal{G}_{i+1/2,j}^{n}={{G}_{i,j+1/2}}\left( {{\mathbf{W}}^{n}},{{t}_{n}} \right)+\frac{\Delta t}{6}\left[ {{\partial }_{t}}{{G}_{i,j+1/2}}\left( {{\mathbf{W}}^{n}},{{t}_{n}} \right)+2{{\partial }_{t}}{{G}_{i,j+1/2}}\left( {{\mathbf{W}}^{*}},{{t}_{*}} \right) \right].
\end{equation}
Update $\mathbf{W}_{ij}^{n+1}$ by
	 \begin{equation}
\mathbf{W}_{ij}^{n+1}=\mathbf{W}_{ij}^{n}-\frac{\Delta t}{\Delta x}\left[ \mathcal{F}_{i+1/2,j}^{n}-\mathcal{F}_{i-1/2,j}^{n} \right]-\frac{\Delta t}{\Delta y}\left[ \mathcal{G}_{i+1/2,j}^{n}-\mathcal{G}_{i-1/2,j}^{n} \right].
\end{equation}
	In summary, for most of the tests in the third section, the flux function from the distribution function Eq. (\ref{equ10}) is typically used. As analyzed  in~\cite{AEAA}, in smooth region the Euler equation with leading error $\mathcal{O}\left( {{\left( \Delta x \right)}^{5}},{{\left( \Delta t \right)}^{4}} \right)$ is solved in the S2O4 scheme, and the NS equations with the error $\mathcal{O}\left( {{\left( \Delta x \right)}^{5}},\tau \Delta t \right)$. Normally, the condition $\tau \ll \Delta t$ is satisfied for many flow computations.

\subsection{Hybrid WENO-AO method for spatial reconstruction}
\label{sec2.3}
In this following, we present the main idea about the hybrid WENO scheme. We describe how to determine the point-wise values and derivatives required on the interface in the present hybrid WENO scheme. Next, we extend the HKGS with hybrid WENO method to the two-dimensional case (the process for extending to the three-dimensional case is straightforward). Finally, we provide a detailed analysis of why the improved HGKS can reduce computational costs.
\subsubsection{Reconstruction of non-equilibrium states by hybrid WENO-AO}
\label{sec2.3.1}
The Hybrid WENO-AO scheme extends the WENO-AO scheme by introducing a troubled cell indicator. When reconstructing the target variables on large stencils, we first use the troubled cell indicator to determine whether the solution is smooth on the large stencils. If the troubled cell indicator indicates the presence of discontinuities or shocks, we use the WENO-AO reconstruction method after transforming the conservative variables into the characteristic variables. Otherwise, we use the upwind linear reconstruction method directly. The detailed process is described as follows:\\
{\bfseries Step 1.} A quadratic polynomial is constructed by using least-square method to approximate target variables on the large stencil ${{S}_{3}}=\left\{ {{I}_{i-2}},{{I}_{i-1}},{{I}_{i}},{{I}_{i+1}},{{I}_{i+2}} \right\}$. The quadratic polynomial 
\begin{equation}
\bar{p}_{3}^{r2}\left( x \right)={{\bar{a}}_{0}}+{{\bar{a}}_{1}}x+{{\bar{a}}_{2}}{{x}^{2}}
\end{equation}
is constructed by requiring\\
$\frac{1}{\Delta x}\int_{{{I}_{i+j}}}{\bar{p}_{3}^{r2}\left( x \right)dx}={{\bar{Q}}_{i+j}},j=-2,-1,0,1,2$ and $\min \left\{ {{\sum\limits_{j=i-2,j\ne i}^{i+2}{\left[ \frac{1}{\Delta x}\int_{{{I}_{j}}}{\bar{p}_{3}^{r2}\left( x \right)-{{{\bar{Q}}}_{j}}dx} \right]}}^{2}} \right\}$.\\
The explicit formulas for ${{\bar{a}}_{1}}$ and ${{\bar{a}}_{2}}$ are
\begin{equation}
\left\{ \begin{aligned}
  & {{{\bar{a}}}_{1}}=-\frac{2{{{\bar{Q}}}_{i-2}}+{{{\bar{Q}}}_{i-1}}-{{{\bar{Q}}}_{i+1}}-2{{{\bar{Q}}}_{i+2}}}{10\Delta x} \\ 
 & {{{\bar{a}}}_{2}}=\frac{4{{{\bar{Q}}}_{i-2}}+{{{\bar{Q}}}_{i-1}}-10{{{\bar{Q}}}_{i}}+{{{\bar{Q}}}_{i+1}}+4{{{\bar{Q}}}_{i+2}}}{34\Delta {{x}^{2}}} \\ 
\end{aligned} \right..
\end{equation}
If second-order derivative $\left| {{\left( \bar{p}_{3}^{r2}\left( x \right) \right)}^{\prime \prime }} \right|=\left| 2{{{\bar{a}}}_{2}} \right|\le {{\zeta }_{1}}$, the cell ${{I}_{i}}$ is judged as a non-troubled cell, otherwise, go to Step 2. ${{\zeta }_{1}}=5/17\Delta x$ is usually used for fifth-order method.\\
{\bfseries Step 2.} The extreme point of $\bar{p}_{3}^{r2}\left( x \right)$ is calculated as $\bar{x}=-{{\bar{a}}_{1}}/2{{\bar{a}}_{2}}$. If $\bar{x}\notin {{S}_{3}}=\left[ -\frac{5}{2}\Delta x,\ \frac{5}{2}\Delta x \right]$, the cell ${{I}_{i}}$ is judged as a non-troubled cell, otherwise, go to Step 3.\\
{\bfseries Step 3.} A new quadratic polynomial is constructed in a conservative way to approximate target variables on the large stencil ${{S}_{1}}=\left\{ {{I}_{i-1}},{{I}_{i}},{{I}_{i+1}} \right\}$ which is
\begin{equation}
\hat{p}_{3}^{r2}\left( x \right)={{\hat{a}}_{0}}+{{\hat{a}}_{1}}x+{{\hat{a}}_{2}}{{x}^{2}}.
\end{equation}
A point is determined as $\hat{x}=-{{\hat{a}}_{1}}/\left\{ 2{{{\hat{a}}}_{2}}+\left[ 1-sign\left( \left| {{{\hat{a}}}_{2}} \right| \right) \right]{{10}^{-30}} \right\}$, where $2{{\hat{a}}_{2}}+\left[ 1-sign\left( \left| {{{\hat{a}}}_{2}} \right| \right) \right]{{10}^{-30}}$ is just used to avid the denominator being zero. If $\left| \bar{x}-\hat{x} \right|\le {{\zeta }_{2}}$ where ${{\zeta }_{2}}=\Delta x/4$ is usually used for fifth-order method, the cell ${{I}_{i}}$ is judged as a non-troubled cell, otherwise it is judged as a troubled cell, go to Step 4.\\
{\bfseries Step 4.} If ${{I}_{i}}$ is judged as a troubled cell, the cells $\left\{ {{I}_{i-1}},\ {{I}_{i+1}} \right\}$ are also judged as troubled cells for the fifth order method. More details about the troubled cell indicator of the hybrid WENO-AO method are presented in ~\cite{ODHW}.\\
{\bfseries Step 5.} If ${{I}_{i}}$ isn’t judged as a troubled cell, the upwind linear reconstruction will be applied to the target variables in the way of component by component. On the large stencil ${{S}_{3}}=\left\{ {{S}_{0}},{{S}_{1}},{{S}_{2}} \right\}$, the point values and derivatives can at cell interface ${{x}_{i+1/2}}$ can be written as 
 \begin{equation}
Q_{i+1/2}^{l}=\frac{1}{60}\left( 2{{{\bar{Q}}}_{i-2}}-13{{{\bar{Q}}}_{i-1}}+47{{{\bar{Q}}}_{i}}+27{{{\bar{Q}}}_{1}}-3{{{\bar{Q}}}_{i+2}} \right)
\end{equation}
\begin{equation}
{{\left( Q_{x}^{l} \right)}_{i+1/2}}=\frac{1}{12\Delta x}\left( {{{\bar{Q}}}_{i-1}}-15{{{\bar{Q}}}_{i}}+15{{{\bar{Q}}}_{1}}-{{{\bar{Q}}}_{i+2}} \right).
\end{equation}
The point values and derivatives can at cell interface ${{x}_{i-1/2}}$ can be written as
\begin{equation}
Q_{i-1/2}^{r}=\frac{1}{60}\left( -3{{{\bar{Q}}}_{i-2}}+27{{{\bar{Q}}}_{i-1}}+47{{{\bar{Q}}}_{i}}-13{{{\bar{Q}}}_{1}}+2{{{\bar{Q}}}_{i+2}} \right)
\end{equation}
\begin{equation}
{{\left( Q_{x}^{r} \right)}_{i-1/2}}=\frac{1}{12\Delta x}\left( {{{\bar{Q}}}_{i-2}}-15{{{\bar{Q}}}_{i-1}}+15{{{\bar{Q}}}_{0}}-{{{\bar{Q}}}_{i+1}} \right).
\end{equation}
Therefore, the upwind linear reconstruction procedure is effective and concise in the smooth region. \par
If ${{I}_{i}}$ is judged as a troubled cell, the reconstruction performed for the characteristic variables to eliminate the spurious oscillation and improve the stability. The $Q$ are defined as conservative variables. The $U={{R}^{-1}}Q$ are defined as characteristic variables, where $R$ is the right eigenmatrix of Jacobian matrix ${{n}_{x}}\left( \partial F/\partial Q \right)+{{n}_{y}}\left( \partial G/\partial Q \right)$. Here $Q$ approximately are the averaged conservative variables from both sides of the cell interface. With the reconstructed values, the conservative variables can be obtained by the inverse projection. And then the WENO-AO method is adopted in the reconstruction procedure. The detail formulation of fifth order WENO-AO method is the following. \par
 Three sub-stencils
 \begin{equation}
{{S}_{0}}=\left\{ {{I}_{i-2}},{{I}_{i-1}},{{I}_{i}} \right\},{{S}_{1}}=\left\{ {{I}_{i-1}},{{I}_{i}},{{I}_{i+1}} \right\},{{S}_{1}}=\left\{ {{I}_{i}},{{I}_{i+1}},{{I}_{i+2}} \right\}
\end{equation}
are selected to reconstruct the left interface value $Q_{i+1/2}^{l}$ and derivative ${{\left( Q_{x}^{l} \right)}_{i+1/2}}$ at the cell interface ${{x}_{i+1/2}}$ and the right interface value $Q_{i-1/2}^{r}$ and derivative ${{\left( Q_{x}^{r} \right)}_{i-1/2}}$ at the cell interface ${{x}_{i-1/2}}$. The three quadratic polynomials $p_{k}^{r3}\left( x \right)$ corresponding to the sub-stencils ${{S}_{k}},k=0,1,2$ are constructed by requiring
\begin{equation}
\frac{1}{\Delta x}\int_{{{I}_{i-j-k-1}}}{p_{k}^{r3}\left( x \right)dx}={{\bar{Q}}_{i-j-k-1}},j=-1,0,1,
\end{equation}
where $\bar{Q}$ represents the cell-averaged value. Each sub-stencil can achieve a third-order spatial accuracy in $r=3$ smooth case. For the reconstructed polynomials, the point values at the cell interface  ${{x}_{i+1/2}}$ and ${{x}_{i-1/2}}$ is given in terms of the cell-averaged value as follows
\begin{equation}
\begin{aligned}
  & p_{0}^{r3}\left( {{x}_{i+1/2}} \right)=\frac{1}{3}{{{\bar{Q}}}_{i-2}}-\frac{7}{6}{{{\bar{Q}}}_{i-1}}+\frac{11}{6}{{{\bar{Q}}}_{i}},p_{0}^{r3}\left( {{x}_{i-1/2}} \right)=-\frac{1}{6}{{{\bar{Q}}}_{i-2}}+\frac{5}{6}{{{\bar{Q}}}_{i-1}}+\frac{1}{3}{{{\bar{Q}}}_{i}}, \\ 
 & p_{1}^{r3}\left( {{x}_{i+1/2}} \right)=-\frac{1}{6}{{{\bar{Q}}}_{i-1}}+\frac{5}{6}{{{\bar{Q}}}_{i}}+\frac{1}{3}{{{\bar{Q}}}_{i+1}},p_{1}^{r3}\left( {{x}_{i-1/2}} \right)=\frac{1}{3}{{{\bar{Q}}}_{i-1}}+\frac{5}{6}{{{\bar{Q}}}_{i}}-\frac{1}{6}{{{\bar{Q}}}_{i+1}}, \\ 
 & p_{2}^{r3}\left( {{x}_{i+1/2}} \right)=\frac{1}{3}{{{\bar{Q}}}_{i}}+\frac{5}{6}{{{\bar{Q}}}_{i+1}}-\frac{1}{3}{{{\bar{Q}}}_{i+2}},p_{2}^{r3}\left( {{x}_{i-1/2}} \right)=\frac{11}{6}{{{\bar{Q}}}_{i}}-\frac{7}{6}{{{\bar{Q}}}_{i+1}}+\frac{1}{3}{{{\bar{Q}}}_{i+2}}. \\ 
\end{aligned}
\end{equation}\par
	On the large stencil ${{S}_{3}}=\left\{ {{S}_{0}},{{S}_{1}},{{S}_{2}} \right\}$, a fourth-order polynomial $p_{3}^{r5}\left( x \right)$ can be constructed by requiring
\begin{equation}
\frac{1}{\Delta x}\int_{{{I}_{i+j}}}{p_{3}^{r5}\left( x \right)dx}={{\bar{Q}}_{i+j}},j=-2,-1,0,1,2.
\end{equation}
The $p_{3}^{r5}\left( x \right)$ can be written by Balsara as
 \begin{equation}
p_{3}^{r5}\left( x \right)={{\gamma }_{3}}\left( \frac{1}{{{\gamma }_{3}}}p_{3}^{r5}\left( x \right)-\sum\limits_{0}^{2}{\frac{{{\gamma }_{k}}}{{{\gamma }_{3}}}p_{k}^{r3}\left( x \right)} \right)+\sum\limits_{0}^{2}{{{\gamma }_{k}}p_{k}^{r3}\left( x \right)},{{r}_{1}}\ne 0,
\end{equation}
where ${{r}_{k}},l=1,2,3$ are defined linear weights. The linear weights for the large stencil ${{S}_{3}}$ and the sub-stencils ${{S}_{0}}$,${{S}_{1}}$ and ${{S}_{2}}$ are given by
\begin{equation}
{{\gamma }_{3}}={{\gamma }_{Hi}},{{\gamma }_{1}}={{\gamma }_{3}}=\left( 1-{{\gamma }_{Hi}} \right)\left( 1-{{\gamma }_{Lo}} \right)/2,{{\gamma }_{2}}=\left( 1-{{\gamma }_{Hi}} \right){{\gamma }_{Lo}},
\end{equation}
which satisfy $\sum\limits_{0}^{3}{{{\gamma }_{k}}}=1$. Typically, ${{\gamma }_{Hi}}\in \left[ 0.85,\ 0.95 \right]$ and ${{\gamma }_{Lo}}\in \left[ 0.85,\ 0.95 \right]$. These linear weights are more flexible than WENO-Z type linear weights. And when the smoothness indicators shows that the large stencil ${{S}_{3}}$ is smooth, the 5th order accuracy can be guaranteed. The smoothness indicators shows that the third-order accuracy can be guaranteed even when it is not smooth. The point value at the cell interface ${{x}_{i+1/2}}$ and ${{x}_{i-1/2}}$ can be given as
\begin{equation}
\begin{aligned}
  & p_{3}^{r5}\left( {{x}_{i+1/2}} \right)=\frac{1}{60}\left( 2{{{\bar{Q}}}_{i-2}}-13{{{\bar{Q}}}_{i-1}}+47{{{\bar{Q}}}_{i}}+27{{{\bar{Q}}}_{i+1}}-3{{{\bar{Q}}}_{i+2}} \right), \\ 
 & p_{3}^{r5}\left( {{x}_{i-1/2}} \right)=\frac{1}{60}\left( -3{{{\bar{Q}}}_{i-2}}+27{{{\bar{Q}}}_{i-1}}+47{{{\bar{Q}}}_{i}}-13{{{\bar{Q}}}_{i+1}}+2{{{\bar{Q}}}_{i+2}} \right). \\ 
\end{aligned}
\end{equation}\par
To deal with discontinuities and avoid the loss of order of accuracy at inflection points, the non-normalized WENO-Z type nonlinear weights ~\cite{AIWE} are introduced as follows
 \begin{equation}
{{\omega }_{k}}={{\gamma }_{k}}\left( 1+\frac{\delta }{{{\beta }_{k}}+\varepsilon } \right),
\end{equation}
where the global smooth indicator $\delta $ is designed as
\begin{equation}
\delta =\frac{1}{3}\left( \left| \beta _{3}^{r5}-\beta _{0}^{r3} \right|+\left| \beta _{3}^{r5}-\beta _{1}^{r3} \right|+\left| \beta _{3}^{r5}-\beta _{3}^{r3} \right| \right)=O\left( \Delta {{x}^{4}} \right).
\end{equation}
The smoothness indicators ${{\beta }_{k}}$ are defined as 
\begin{equation}
{{\beta }_{k}}=\sum\limits_{q=1}^{{{q}_{k}}}{\Delta {{x}^{2q-1}}}\int_{{{x}_{i-1/2}}}^{{{x}_{i+1/2}}}{{{\left( \frac{{{d}^{q}}}{d{{x}^{q}}}{{p}_{k}}\left( x \right) \right)}^{2}}dx}=O\left( \Delta {{x}^{2}} \right),
\end{equation}
where ${{q}_{k}}$ is the order of ${{p}_{k}}\left( x \right)$. For $p_{k}^{r3},k=0,1,2,{{q}_{k}}=2$; for  $p_{3}^{r5},{{q}_{3}}=4$. $\varepsilon ={{10}^{-8}}$ is taken in current work.\par
The normalized weights are given by
 \begin{equation}
{{\bar{\omega }}_{k}}=\frac{{{\omega }_{k}}}{\sum\limits_{0}^{3}{{{\omega }_{k}}}}.
\end{equation}
Then the final form of the reconstructed polynomial is
 \begin{equation}
{{p}^{AO\left( 5,3 \right)}}\left( x \right)={{\bar{\omega }}_{3}}\left( \frac{1}{{{\gamma }_{3}}}p_{3}^{r5}\left( x \right)-\sum\limits_{0}^{2}{\frac{{{\gamma }_{k}}}{{{\gamma }_{3}}}p_{k}^{r3}\left( x \right)} \right)+\sum\limits_{0}^{2}{{{{\bar{\omega }}}_{k}}p_{k}^{r3}\left( x \right)}.
\end{equation}
The desired values at the cell interfaces can be fully written as 
 \begin{equation}
Q_{i-1/2}^{r}={{P}^{AO\left( 5,3 \right)}}\left( {{x}_{i-1/2}} \right),Q_{i+1/2}^{l}={{P}^{AO\left( 5,3 \right)}}\left( {{x}_{i+1/2}} \right).
\end{equation}
The WENO-AO reconstruction procedure is over because it is applied to schemes with Riemann solvers where only point-wise values are needed. In order to calculate the GKS flux, we supplement the derivatives at the cell interfaces on the large stencil and sub-stencils as follows
 \begin{equation}
\begin{aligned}
  & \left( {{p}_{x}} \right)_{0}^{r3}\left( {{x}_{i+1/2}} \right)=\left( {{{\bar{Q}}}_{i-2}}-3{{{\bar{Q}}}_{i-1}}+2{{{\bar{Q}}}_{i}} \right)/\Delta x,\left( {{p}_{x}} \right)_{0}^{r3}\left( {{x}_{i-1/2}} \right)=\left( -{{{\bar{Q}}}_{i-1}}+{{{\bar{Q}}}_{i}} \right)/\Delta x, \\ 
 & \left( {{p}_{x}} \right)_{1}^{r3}\left( {{x}_{i+1/2}} \right)=\left( -{{{\bar{Q}}}_{i}}+{{{\bar{Q}}}_{i+1}} \right)/\Delta x,\left( {{p}_{x}} \right)_{1}^{r3}\left( {{x}_{i-1/2}} \right)=\left( -{{{\bar{Q}}}_{i-1}}+{{{\bar{Q}}}_{i}} \right)/\Delta x, \\ 
 & \left( {{p}_{x}} \right)_{2}^{r3}\left( {{x}_{i+1/2}} \right)=\left( -{{{\bar{Q}}}_{i}}+{{{\bar{Q}}}_{i+1}} \right)/\Delta x,\left( {{p}_{x}} \right)_{2}^{r3}\left( {{x}_{i-1/2}} \right)=\left( -2{{{\bar{Q}}}_{i}}+3{{{\bar{Q}}}_{i+1}}-{{{\bar{Q}}}_{i+2}} \right)/\Delta x. \\ 
\end{aligned}
\end{equation}
\begin{equation}
\begin{aligned}
  & \left( {{p}_{x}} \right)_{3}^{r5}\left( {{x}_{i+1/2}} \right)=\frac{1}{12\Delta x}\left( {{{\bar{Q}}}_{i-1}}-15{{{\bar{Q}}}_{i}}+15{{{\bar{Q}}}_{i+1}}-{{{\bar{Q}}}_{i+2}} \right), \\ 
 & \left( {{p}_{x}} \right)_{3}^{r5}\left( {{x}_{i-1/2}} \right)=\frac{1}{12\Delta x}\left( {{{\bar{Q}}}_{i-2}}-15{{{\bar{Q}}}_{i-1}}+15{{{\bar{Q}}}_{i}}-{{{\bar{Q}}}_{i+1}} \right). \\ 
\end{aligned}
\end{equation}
The desired derivatives at the cell interfaces can be fully determined as
\begin{equation}
{{\left( Q_{x}^{r} \right)}_{i-1/2}}=P_{x}^{^{AO\left( 5,3 \right)}}\left( {{x}_{i-1/2}} \right),{{\left( Q_{x}^{l} \right)}_{i+1/2}}=P_{x}^{^{AO\left( 5,3 \right)}}\left( {{x}_{i+1/2}} \right).
\end{equation}
{\bfseries Remark 1.} In the HGKS with the hybrid WENO-AO method described above, both upwind linear reconstruction and WENO-AO reconstruction ensure fifth-order accuracy of the non-equilibrium state. However, WENO-AO reconstruction is more computationally intensive than upwind linear reconstruction due to its higher complexity. This is because the computational cost of the nonlinear weights and high order smooth indicators of the WENO-AO method is already greater than that of WENO-Z or WENO-JS. To reduce computational costs, a troubled cell indicator is employed to identify extreme points on the large stencils that require WENO-AO reconstruction. By doing so, the proportion of WENO-AO reconstruction can be reduced, while maintaining the same level of accuracy.
 
\subsubsection{Reconstruction of equilibrium states}
\label{sec2.3.2}
The non-equilibrium states are obtained by either the upwind linear reconstruction or the WENO-AO reconstruction, and the equilibrium states can be obtained by the following simple method. 
\begin{equation}\label{equ46}
\int{\psi }{{g}^{c}}d\Xi ={{\mathbf{W}}^{c}}=\int_{u>0}{\psi {{g}^{l}}d\Xi }+\int_{u>0}{\psi {{g}^{r}}d\Xi },
\end{equation}
\begin{equation}\label{equ47}
\int{\psi }g_{x}^{c}d\Xi =\mathbf{W}_{x}^{c}=\int_{u>0}{\psi g_{x}^{l}d\Xi }+\int_{u>0}{\psi g_{x}^{r}d\Xi },
\end{equation}
where ${{g}^{c}},g_{x}^{c},g_{xx}^{c}\ldots $ are the equilibrium states and ${{g}^{l,r}},g_{x}^{l,r},g_{xx}^{l,r}\ldots $ are the non-equilibrium states. This is a kinetic-based weighting of the values and derivatives on the left and right sides of the cell interface, while introducing the upwind mechanics. Arithmetic averaging can also be used for smooth flow. In classic WENO5-GKS~\cite{AEAA}, an extra linear polynomial reconstruction for the equilibrium states is required. This method requires no additional process and achieves a 5th-order spatial accuracy for the equilibrium states. In the above way, all components of the microscopic slopes across the interface have been obtained.

\subsubsection{Two-dimensional Reconstruction procedure}
\label{sec2.3.3}
Similar to the HGKS with WENO-AO shown in~\cite{PEFH}. In fact, every step in the reconstruction process is to replace the WENO-AO method with the hybrid WENO-AO method in the new scheme.\par
In the two-dimensional case, the values of each Gaussian point $\left( {{x}_{i+1/2}},{{y}_{jm}} \right),m=1,2$ which need to be obtained by reconstruction are
 $${{W}^{l}},W_{x}^{l},W_{y}^{l},{{W}^{r}},W_{x}^{r},W_{y}^{r}.$$
The scheme for obtaining the values of Gaussian points is as follows through dimension-by-dimensional reconstruction as follows. The time level is omitted here.\\
{\bfseries Step 1.} According to the one-dimensional Hybrid WENO-AO reconstruction in Section \ref{sec2.3.1}, the line averaged reconstructed values and slopes
 $${{\left( {{Q}^{l}} \right)}_{i+1/2,j}},{{\left( Q_{x}^{l} \right)}_{i+1/2,j}},{{\left( {{Q}^{r}} \right)}_{i+1/2,j}},{{\left( Q_{x}^{r} \right)}_{i+1/2,j}}$$
can be obtained along the normal direction by using the cell averaged values ${{\left( {\bar{Q}} \right)}_{i+l,j}}$ and ${{\left( {\bar{Q}} \right)}_{i+l+1,j}}$,$l=-2,\ldots ,2$.\\
{\bfseries Step 2.} Again with the one-dimensional Hybrid WENO-AO reconstruction in Section\ref{sec2.3.1}, the values at each Gaussian point
 $${{\left( {{Q}^{l}} \right)}_{i+1/2,jm}},{{\left( Q_{y}^{l} \right)}_{i+1/2,jm}},{{\left( {{Q}^{r}} \right)}_{i+1/2,jm}},{{\left( Q_{y}^{r} \right)}_{i+1/2,jm}}$$
with $y={{y}_{jm}},m=0,1$ can be obtained by using the line averaged values ${{\left( {{Q}^{l}} \right)}_{i+1/2,j+l}},{{\left( {{Q}^{r}} \right)}_{i+1/2,j+l}},l=-2,\ldots ,2$ constructed above. In the same way, the point-wise derivatives at Gaussian point $\left( {{x}_{i+1/2}},{{y}_{jm}} \right),m=1,2$ ${{\left( Q_{x}^{l} \right)}_{i+1/2,jm}},{{\left( Q_{x}^{r} \right)}_{i+1/2,jm}}$ can be constructed by using the above line averaged derivatives ${{\left( Q_{x}^{l} \right)}_{i+1/2,j+l}},{{\left( Q_{x}^{r} \right)}_{i+1/2,j+l}},l=-2,\ldots ,2$ with the hybrid WENO-AO method in Section\ref{sec2.3.1}. The details about tangential reconstruction are described in ~\cite{PEFH}.\\
{\bfseries Step 3.} The quantities related to the non-equilibrium states at each Gaussian point are all obtained. And then the quantities related to the equilibrium states can be obtained by the unified weighting Eq. (\ref{equ46}) and Eq. (\ref{equ47}).\\
{\bfseries Remark 2.} In 1D, the GKS flux on a cell interface can be reconstructed using WENO-AO  only twice. But in 2D, the GKS flux on a cell interface should be reconstructed using WENO-AO up to six times, including two processes in Step 1 and four processes in Step 2. The WENO-AO reconstruction is used many times for flow variables and slopes in high dimensional tangential reconstruction. Therefore, replacing WENO-AO reconstruction with upwind linear reconstruction without local characteristic decompositions when allowed can save computing costs. Minimizing the use of WENO-AO reconstruction and using more upwind linear reconstruction helps reduce computational costs.

\section{Numerical tests}
\label{sec3}
In this section, 1-D and 2-D numerical tests will be presented to validate the HGKS with hybrid WENO-AO reconstruction. The CPU time for the WENO-AO reconstruction procedure and the hybrid WENO-AO reconstruction procedure are obtained with Intel Core i5-9400 CPU @ 2.90GHz.
For the parameters of the HGKS in the follow tests, the ratio of specific heats takes $\gamma =1.4$. For the inviscid flow, the collision time $\tau $ is
$$\tau ={{c}_{1}}\Delta t+{{c}_{2}}\left| \frac{{{p}_{l}}-{{p}_{r}}}{{{p}_{l}}+{{p}_{r}}} \right|\Delta t,$$
where ${{p}_{l}}$ and ${{p}_{r}}$ denote the pressure on the left and right cell interface. Usually ${{c}_{1}}=0.01$ and ${{c}_{2}}=1$ are chosen in the classic HGKS. But, ${{c}_{1}}=0$ can be safely selected for WENO5-AO GKS and hybrid WENO5-AO GKS in most test cases. Without additional explanation, take ${{c}_{1}}=0.01$ and ${{c}_{2}}=1$ here. The pressure jump term in $\tau $ can add artificial dissipation to enlarge the shock thickness to the scale of numerical cell size in the discontinuous region. Besides, it can keep the non-equilibrium dynamics in the shock layer through the kinetic particle transport to mimic the real physical mechanism inside the shock structure. \par
 For the viscous flow~\cite{AEAA}~\cite{AMDO}, the collision time term related to the viscosity coefficient is defined as 
$$\tau =\frac{\mu }{p}+{{c}_{2}}\left| \frac{{{p}_{l}}-{{p}_{r}}}{{{p}_{l}}+{{p}_{r}}} \right|\Delta t,$$
where $\mu $ is the dynamic viscous coefficient and $p$ is the pressure at the cell interface. In smooth viscous flow region, it reduces to $\tau =\frac{\mu }{p}$. The time step is determined by
$$\Delta t={{C}_{CFL}}Min\left( \frac{\Delta x}{\left\| \mathbf{U} \right\|+{{a}_{s}}},\frac{{{\left( \Delta x \right)}^{2}}}{4\nu } \right),$$
where $\left\| \mathbf{U} \right\|$ is the magnitude of velocities, ${{C}_{CFL}}$ is the CFL number, ${{a}_{s}}$ is the sound speed and $\nu =\frac{\mu }{p}$ is the kinematic viscosity coefficient.

\subsection{Accuracy test in 1-D}
\label{sec3.1}
The advection of density perturbation is tested whose initial condition is set as follows
$$\rho \left( x \right)=1+0.2sin\left( \pi x \right),U\left( x \right)=1,p\left( x \right)=1,x\in \left[ 0,2 \right].$$
Both the left and right sides of the test case are periodic boundary conditions. The analytic solution of the advection of density perturbation is
$$\rho \left( x,t \right)=1+0.2sin\left( \pi \left( x-t \right) \right),U\left( x,t \right)=1,p\left( x,t \right)=1,x\in \left[ 0,2 \right].$$
In the computation, a uniform mesh with $N$  points are used. The time step $\Delta t=0.2\Delta x$ is fixed. The collision time $\tau =0$ is set since the flow is smooth and inviscid. Based on the two-stage fourth-order time-marching method, the HGKS with hybrid WENO-AO method is expected to achieve the same fifth-order spatial accuracy and fourth-order temporal accuracy as analyzed in ~\cite{AEAA}. The ${{L}^{1}},{{L}^{2}}$ and ${{L}^{\infty }}$ errors and corresponding orders at $t=2.0$ are given in the follow tables. WENO-AO reconstruction is used near the two extreme points of a sine wave, as judged by the troubled cell indicator. The remaining region is reconstructed using upwind linear reconstruction. With the mesh refinement in Table \ref{1Daccuracy01} and Table \ref{1Daccuracy02}, the expected orders of accuracy are obtained and the numerical errors are identical.\\
\begin{table}
\begin{tabular}{c|cc|cc|cc}
	\hline
	mesh length & ${{L}^{1}}$error & Order & ${{L}^{2}}$error & Order & ${{L}^{\infty }}$error & Order\\
	\hline
	1/5 & 9.0941825e-04 &  & 1.0207546e-03 &  & 1.4309840e-03 & \\ 
	1/10 & 2.8778750e-05 & 4.98 & 3.1964978e-05 & 5.00 & 4.7027062e-05 & 4.93 \\
	1/20 & 9.0471847e-07 & 4.99 & 1.0020469e-06 & 5.00 & 1.4850604e-06 & 4.98\\
	1/40 & 2.8269469e-08 & 5.00 & 3.1333382e-08 & 5.00 & 4.6521236e-08 & 5.00\\
	1/80 & 8.8554643e-10 & 5.00 & 9.8172632e-10 & 5.00 & 1.4546295e-09 & 5.00\\
	\hline
\end{tabular}
\centering 
	\caption{\label{1Daccuracy01}Accuracy test in 1-D for the advection of density perturbation by the WENO-AO reconstruction. $\Delta t=0.2\Delta x.$}
\end{table}
\begin{table}
\begin{tabular}{c|cc|cc|cc}
	\hline
	mesh length & ${{L}^{1}}$error & Order & ${{L}^{2}}$error & Order & ${{L}^{\infty }}$error & Order\\
	\hline
	1/5 & 8.8190167e-04 &  & 9.9663892e-04 &  & 1.4171754e-03 & \\ 
	1/10 & 2.8712310e-05 & 4.94 & 3.1905636e-05 & 4.97 & 4.7018003e-05 & 4.91 \\
	1/20 & 9.0460559e-07 & 4.99 & 1.0019354e-06 & 4.99 & 1.4850249e-06 & 4.98\\
	1/40 & 2.8269469e-08 & 5.00 & 3.1333382e-08 & 5.00 & 4.6521236e-08 & 5.00\\
	1/80 & 8.8554523e-10 & 5.00 & 9.8172512e-10 & 5.00 & 1.4546323e-09 & 5.00\\
	\hline
\end{tabular}
\centering 
	\caption{\label{1Daccuracy02}Accuracy test in 1-D for the advection of density perturbation by the hybrid WENO-AO reconstruction. $\Delta t=0.2\Delta x.$}
\end{table}
\subsection{One-dimensional Riemann problems}
\label{sec3.2}
The reference solutions for the following 1-D Riemann problems were obtained using classic WENO5-GKS with 10,000 uniform mesh points. A comparison of the CPU times for spatial reconstruction demonstrates the advantages of the hybrid WENO-AO method for HGKS. For the same test case, the CPU time for the remaining program components is identical between HGKS with WENO-AO and HGKS with hybrid WENO-AO. For instance, in the SOD problem, the CPU time for the rest of the program is approximately 0.196 seconds for both methods. However, the CPU time for the reconstruction procedure differs, as shown in Table \ref{1DCPU}. \\
\begin{table}
\begin{tabular}{c|c|p{4cm}|p{3cm}|c}
	\hline
	Numerical example & Grid points & CPU time of hybrid WENO-AO (s) & CPU time of WENO-AO (s) & Ratio\\
	\hline
	Sod problem & 100 & 0.016 & 0.049 & 32.65\% \\ 
	Shu-Osher problem & 400 & 0.245 & 1.937 &12.65\%  \\
	Blast wave problem & 400 & 0.773 & 3.176 & 23.34\% \\
	\hline
\end{tabular}
\centering 
	\caption{\label{1DCPU}Computing time of the hybrid WENO-AO reconstruction procedure and the WENO-AO reconstruction procedure in 1-D test cases.}
\end{table}
\emph{(a) Sod problem}\par
The computational domain for the Sod problem is $\left[ 0,\ 1 \right]$ with 100 uniform mesh points. The solutions of the Sod problem are presented at $t=0.2$ with non-reflecting boundary condition on both ends. The initial condition is given by
 $$\left( \rho ,U,p \right)=\left\{ \begin{aligned}
  & \left( 1,0,1 \right),0<x<0.5, \\ 
 & \left( 0.125,0,0.1 \right),0.5<x<1. \\ 
\end{aligned} \right.$$

\begin{figure}[htbp]
	\centering
	\subfigure{\includegraphics[width=0.48\textwidth]{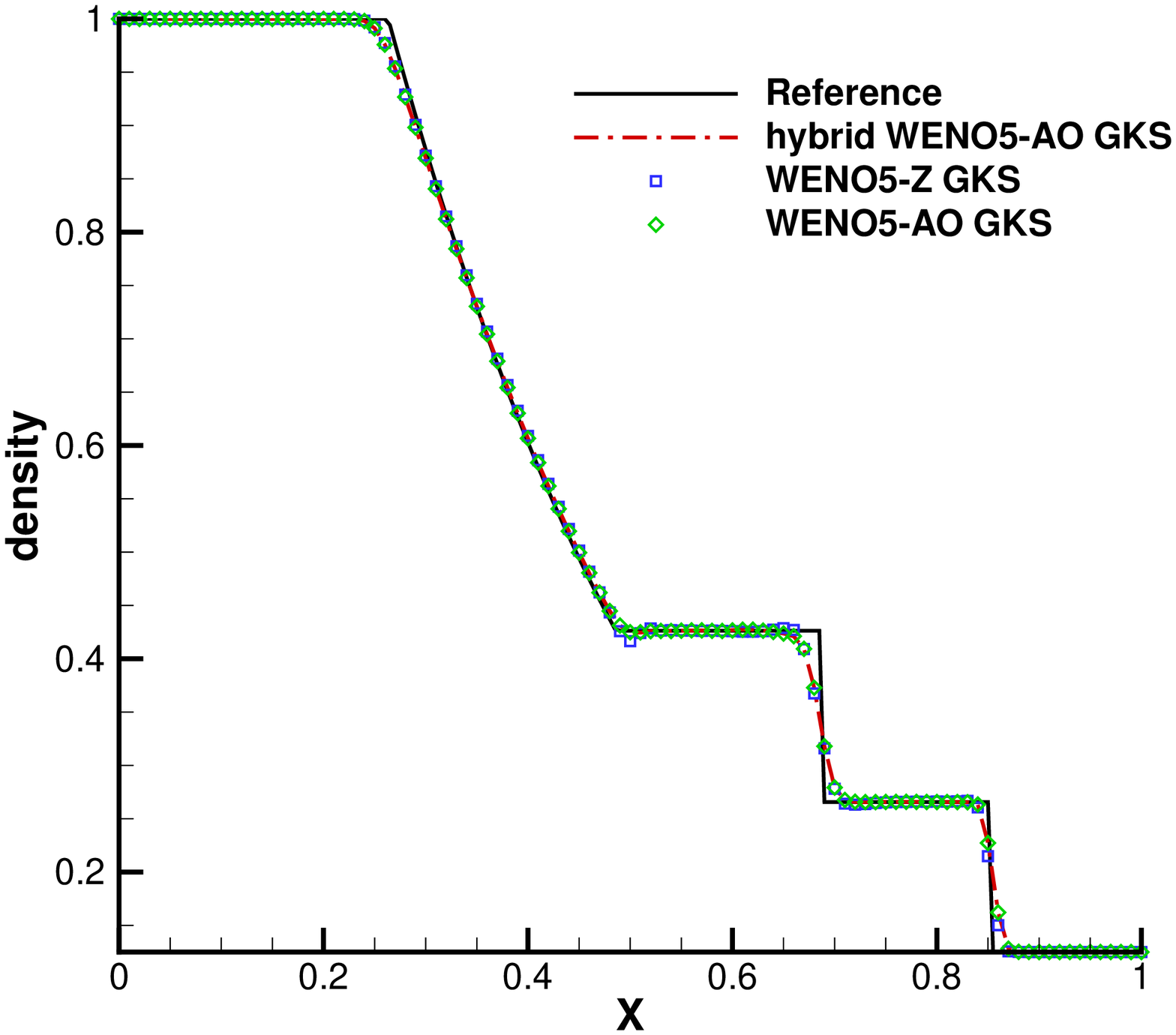}}
	\subfigure{\includegraphics[width=0.48\textwidth]{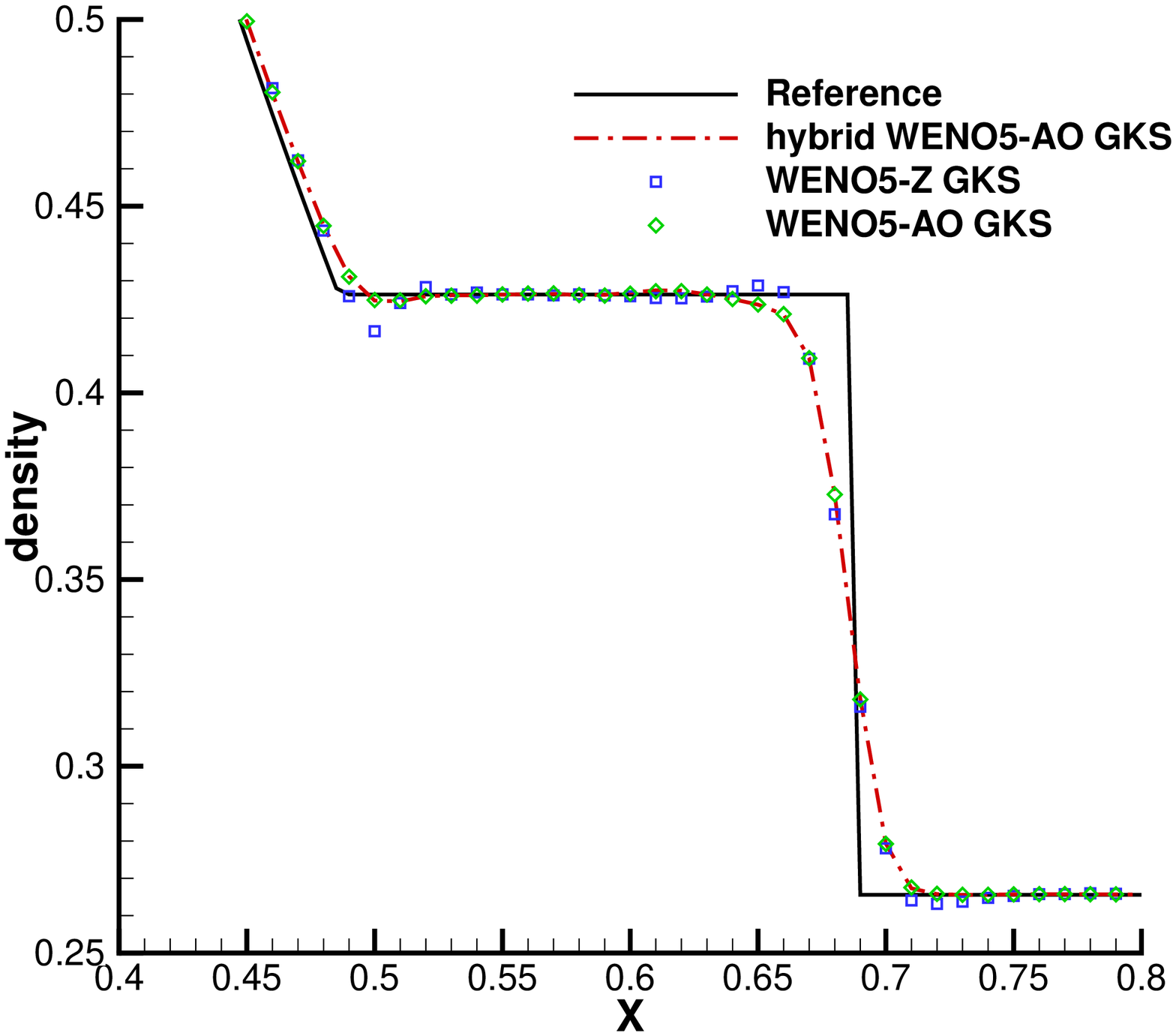}}
	\subfigure{\includegraphics[width=0.48\textwidth]{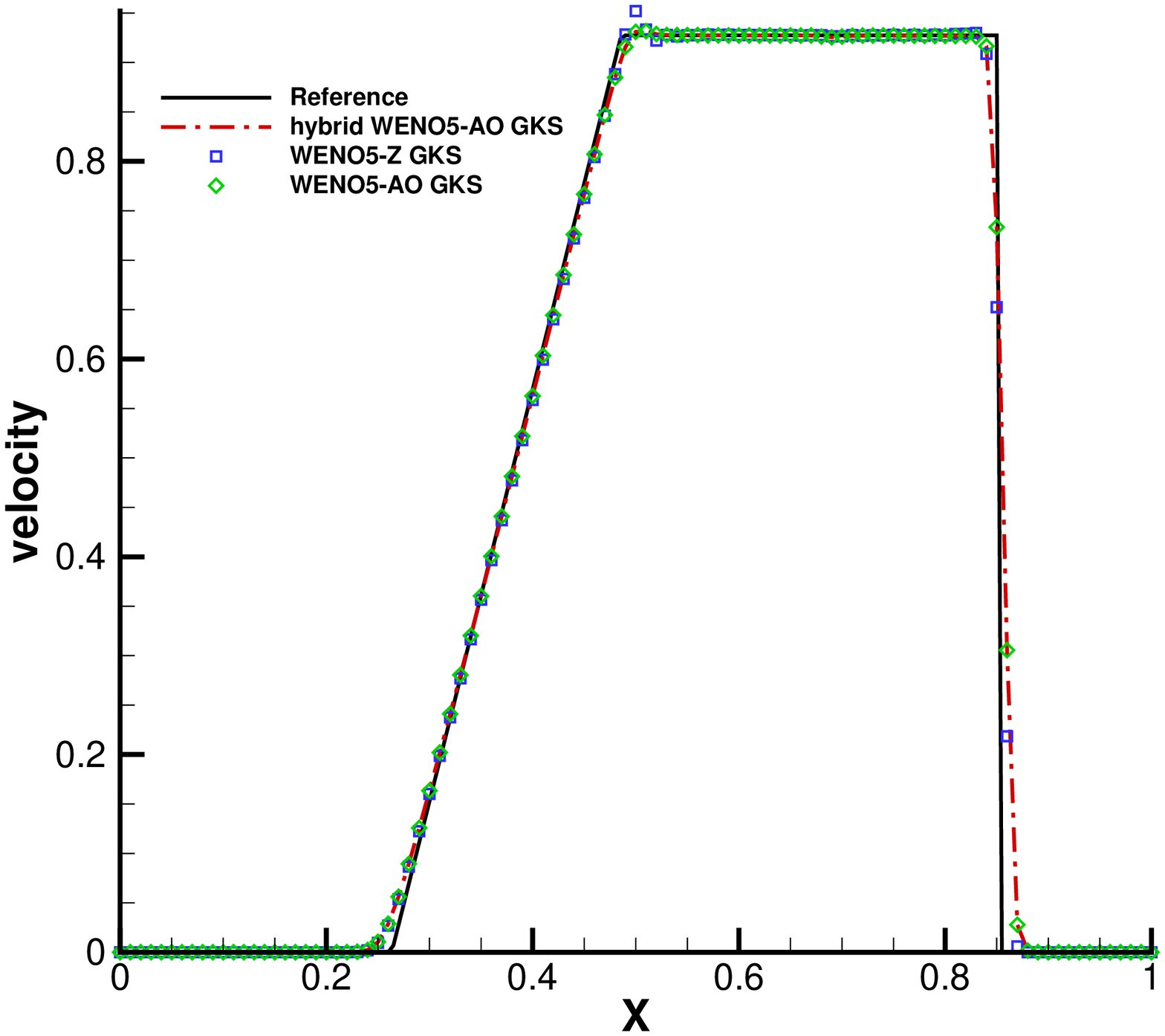}}
	\subfigure{\includegraphics[width=0.48\textwidth]{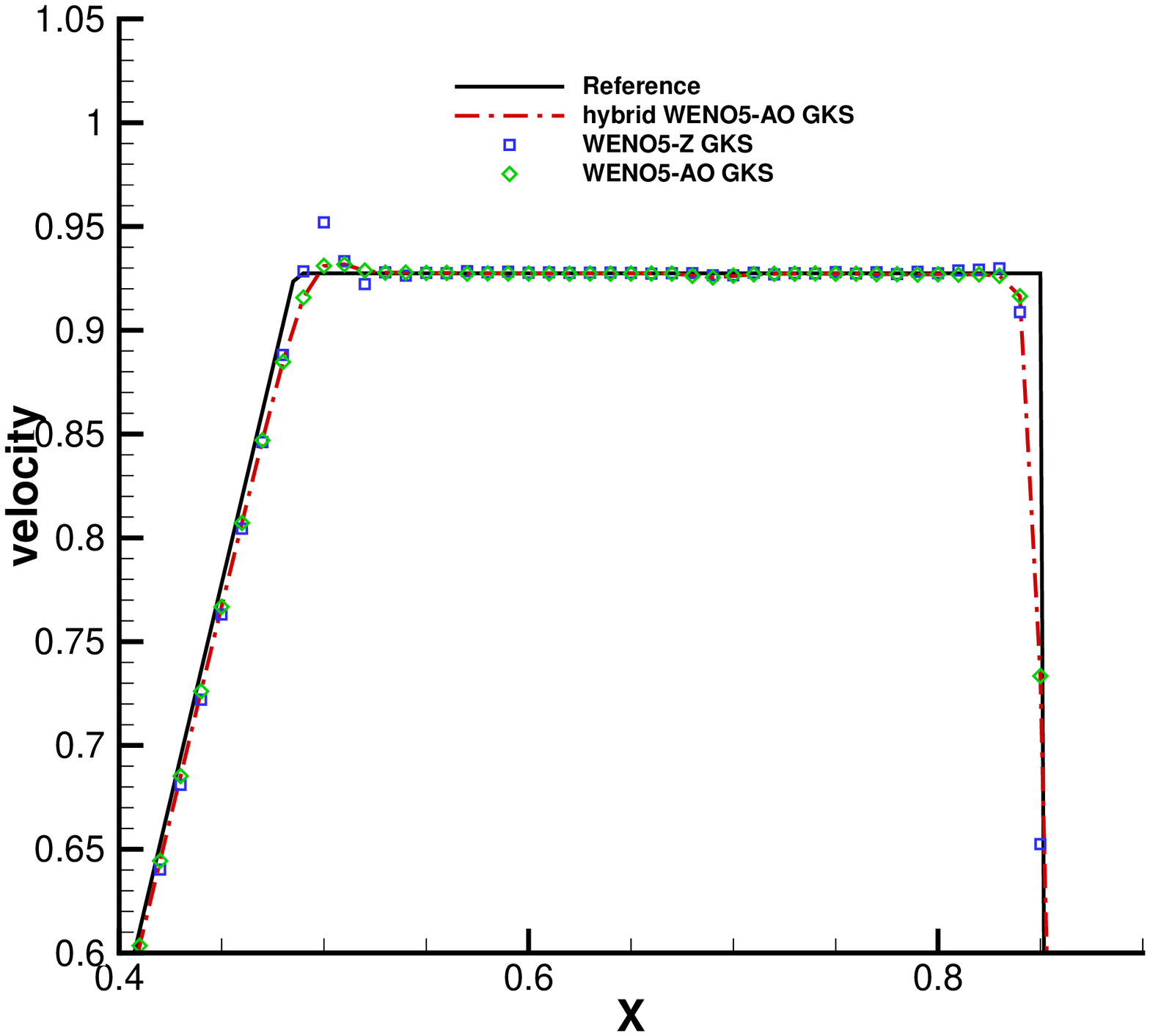}}
	\subfigure{\includegraphics[width=0.48\textwidth]{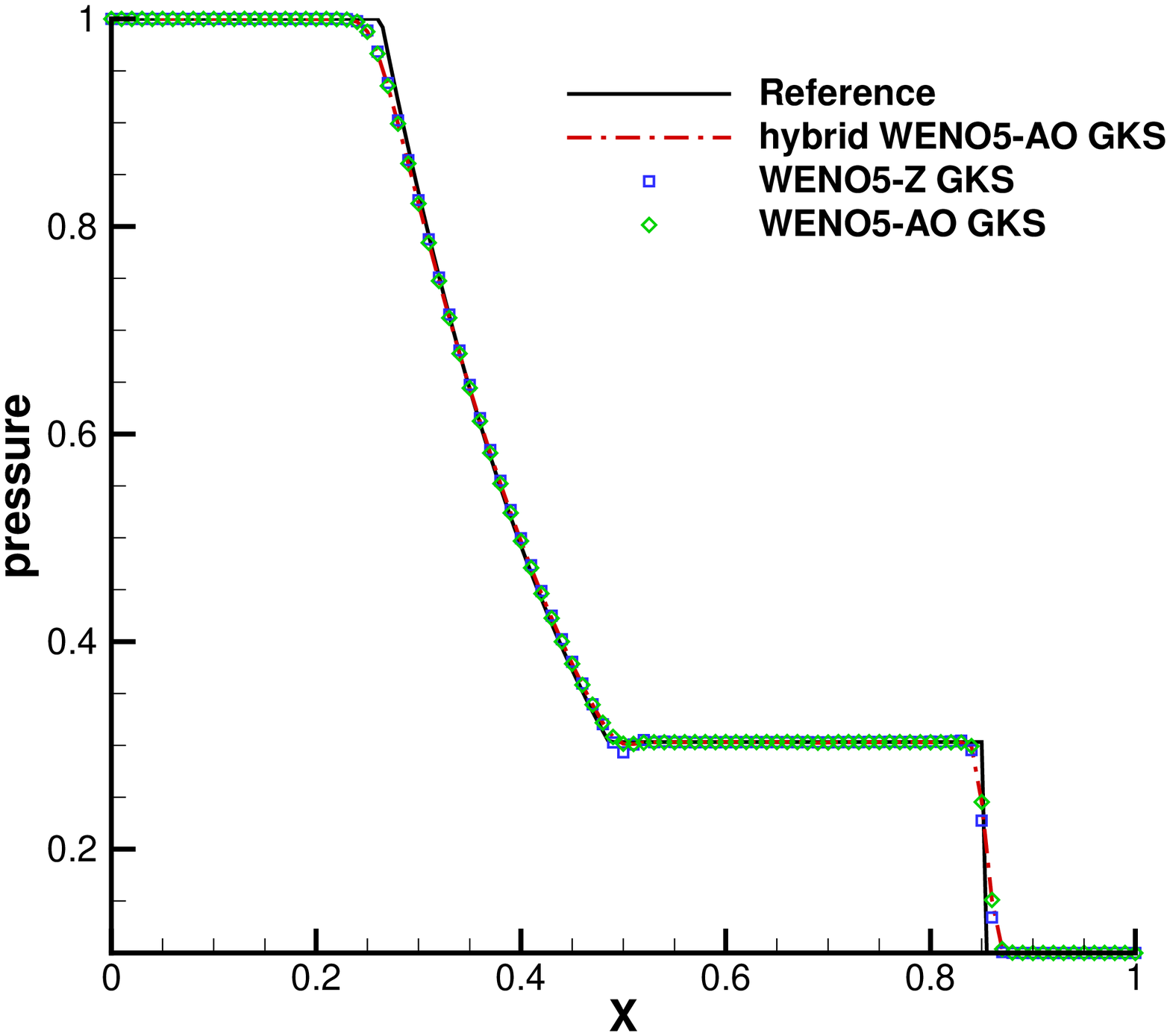}}
	\subfigure{\includegraphics[width=0.48\textwidth]{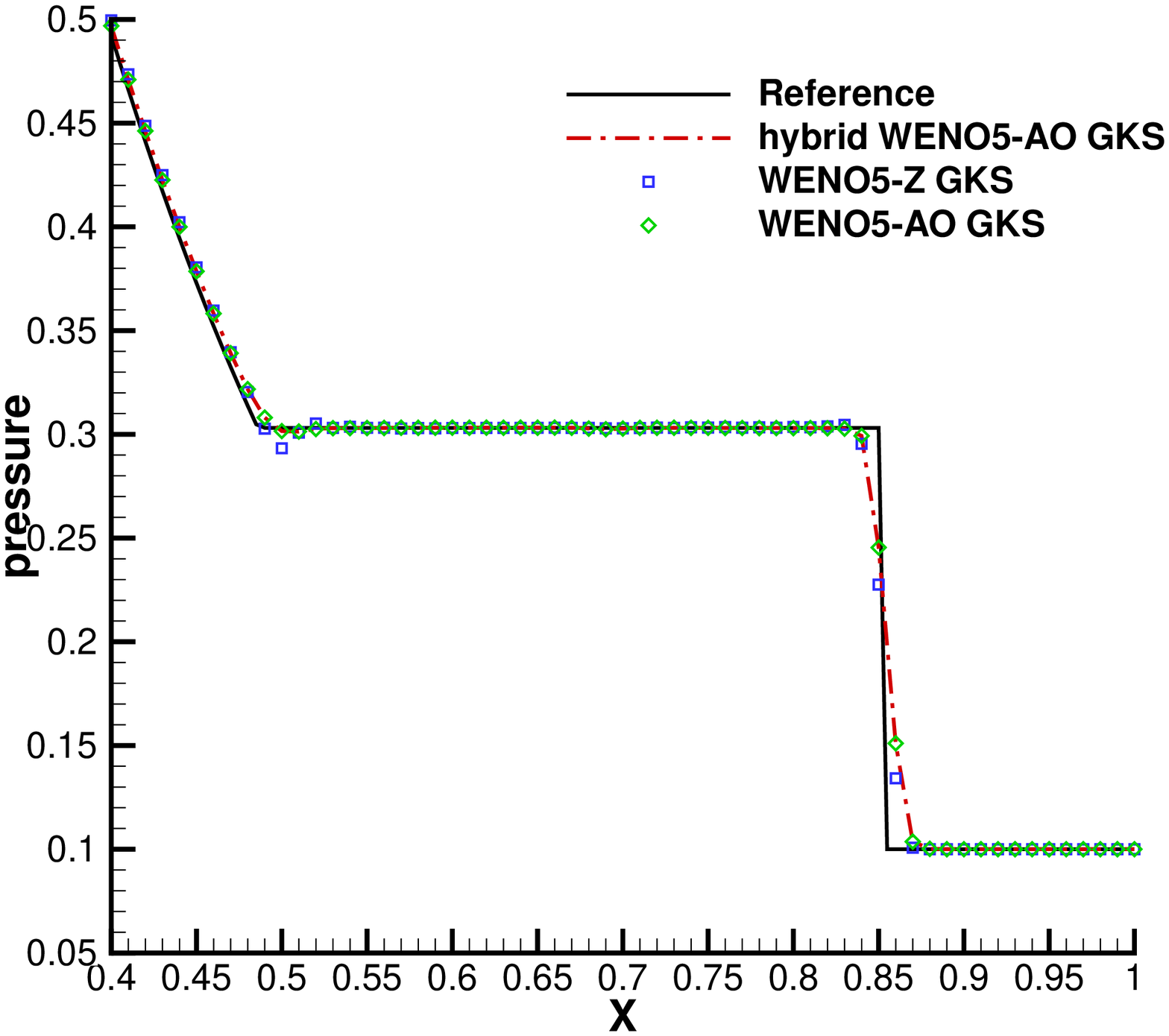}}
	\caption{Sod problem: the density, velocity and pressure distributions and local enlargements with 100 cells. $CFL=0.5.$ $T=0.2.$ }
	\label{figSOD}
\end{figure}

In Fig. \ref{figSOD}, a comparison is made between the results of the class WENO5-Z GKS, WENO5-AO GKS and the hybrid WENO5-AO GKS. Overall, the results of the three reconstruction methods are in good agreement with the reference solutions. From the local enlargements in Fig. \ref{figSOD}, it is observed that the solutions from the classic HGKS with WENO-Z show undershoot or overshoot around the corner of the rarefaction wave. The results obtained using the HGKS with both WENO-AO and the hybrid WENO-AO are almost identical due to the kinetic-weighting method, which is analyzed in Section\ref{sec2.3.2}. It can be seen from Table \ref{1DCPU} that the hybrid WENO-AO reconstruction can save 67.35\% of CPU time compared to WENO-AO reconstruction for the CPU time of the reconstruction procedure of the HGKS in the Sod problem.\\
\emph{(b) Shu-Osher problem}\par
\begin{figure}[htbp]
	\centering
	\subfigure{\includegraphics[width=0.48\textwidth]{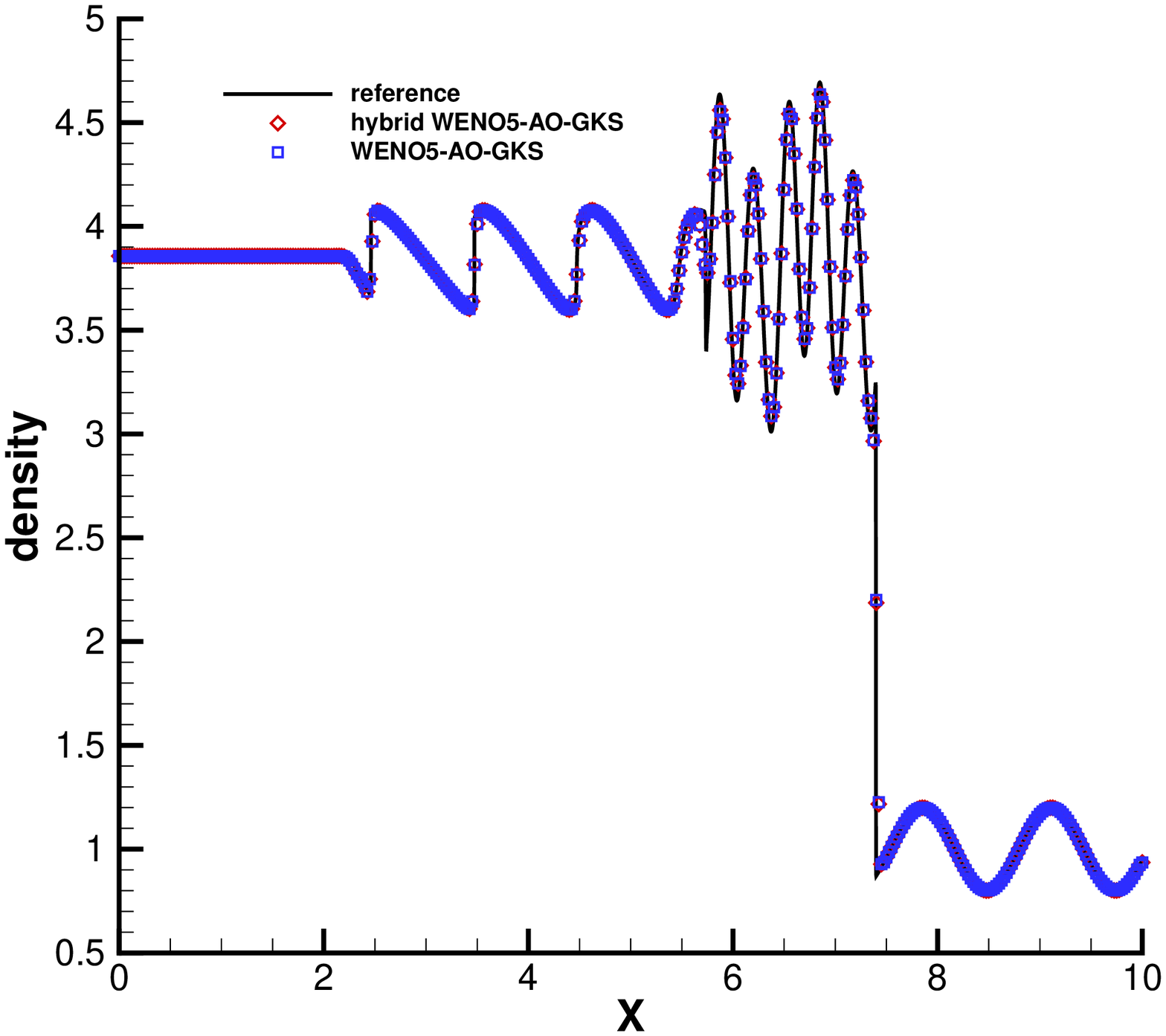}}
	\subfigure{\includegraphics[width=0.48\textwidth]{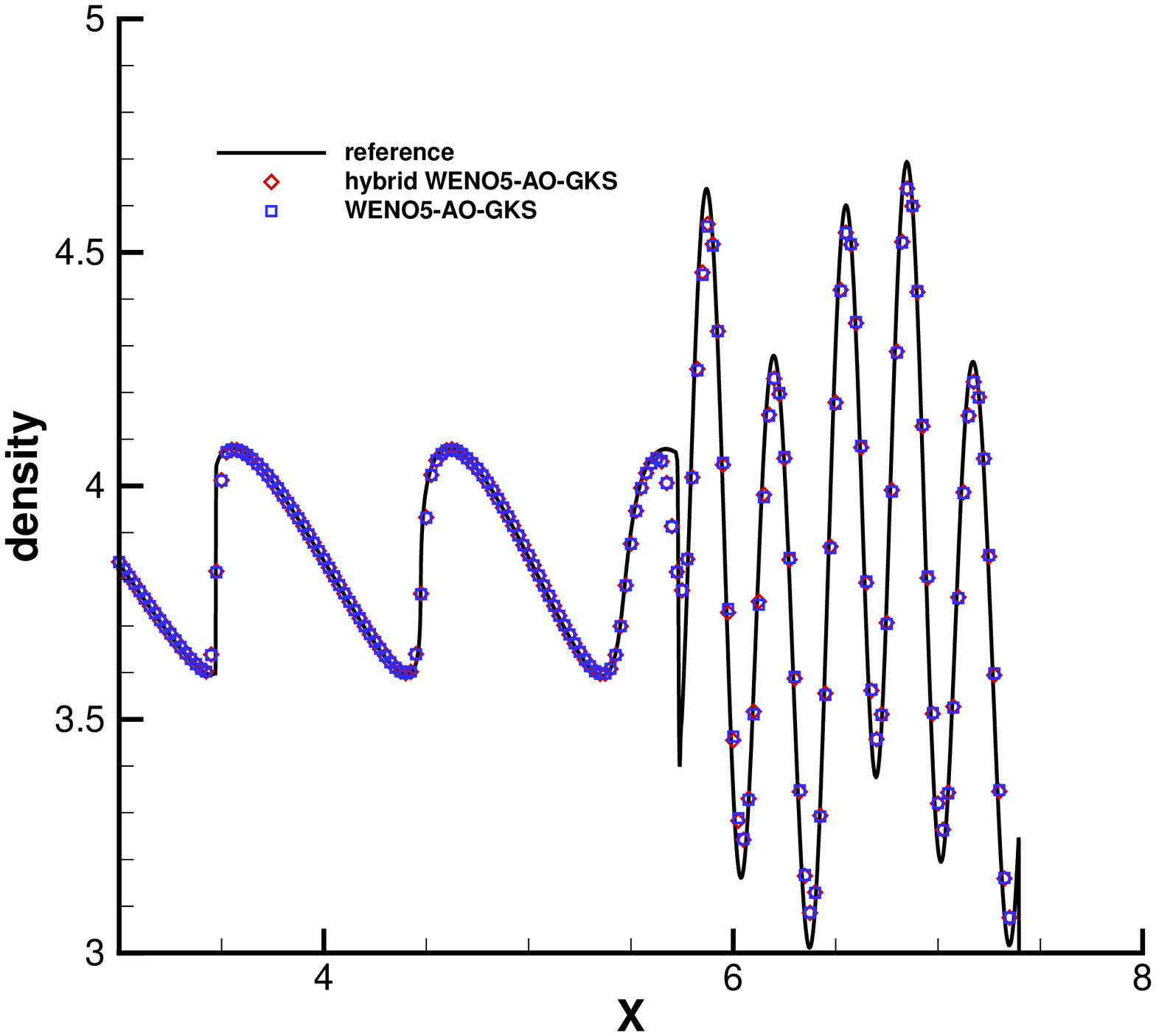}}
	\caption{Shu-Osher problem: the density distributions and local enlargements for WENO-AO reconstruction and hybrid WENO-AO reconstruction with 400 cells. $CFL=0.5.$ $T=1.8.$ }
	\label{figshu}
\end{figure}

The Shu-Osher shock acoustic interaction ~\cite{EIOE} is computed in $[0,\ 10]$ with 400 mesh points. The non-reflecting boundary condition is given on the left, and the fixed wave profile is extended on the right. The initial conditions are
 $$\left( \rho ,U,p \right)=\left\{ \begin{aligned}
  & \left( 3.857134,2.629369,10.33333 \right),0<x\le 1, \\ 
 & \left( 1+0.2\sin \left( 5x \right),0,1 \right),1\le x\le 10. \\ 
\end{aligned} \right.$$
The Fig. \ref{figshu} presents density profiles and enlargements at $t=1.8$. The values of density computed by the HGKS with WENO-AO reconstruction and hybrid WENO-AO reconstruction are nearly identical with each other. But for the reconstruction procedure, the HGKS with hybrid WENO-AO method requires approximately 87.35\% less cpu time compared to the HGKS with WENO-AO method.\\
\emph{(c) Blast wave problem}\par
The third case is the Woodward-Colella blast~\cite{TNSO} wave problem. The initial conditions for the blast wave problem are given as follows
 $$\left( \rho ,U,p \right)=\left\{ \begin{aligned}
  & \left( 1,0,1000 \right),0\le x<0.1, \\ 
 & \left( 1,0,0.01 \right),0.1\le x<0.9, \\ 
 & \left( 1,0,100 \right),0.9\le x\le 1.0. \\ 
\end{aligned} \right.$$

\begin{figure}[htbp]
	\centering
	\subfigure{\includegraphics[width=0.48\textwidth]{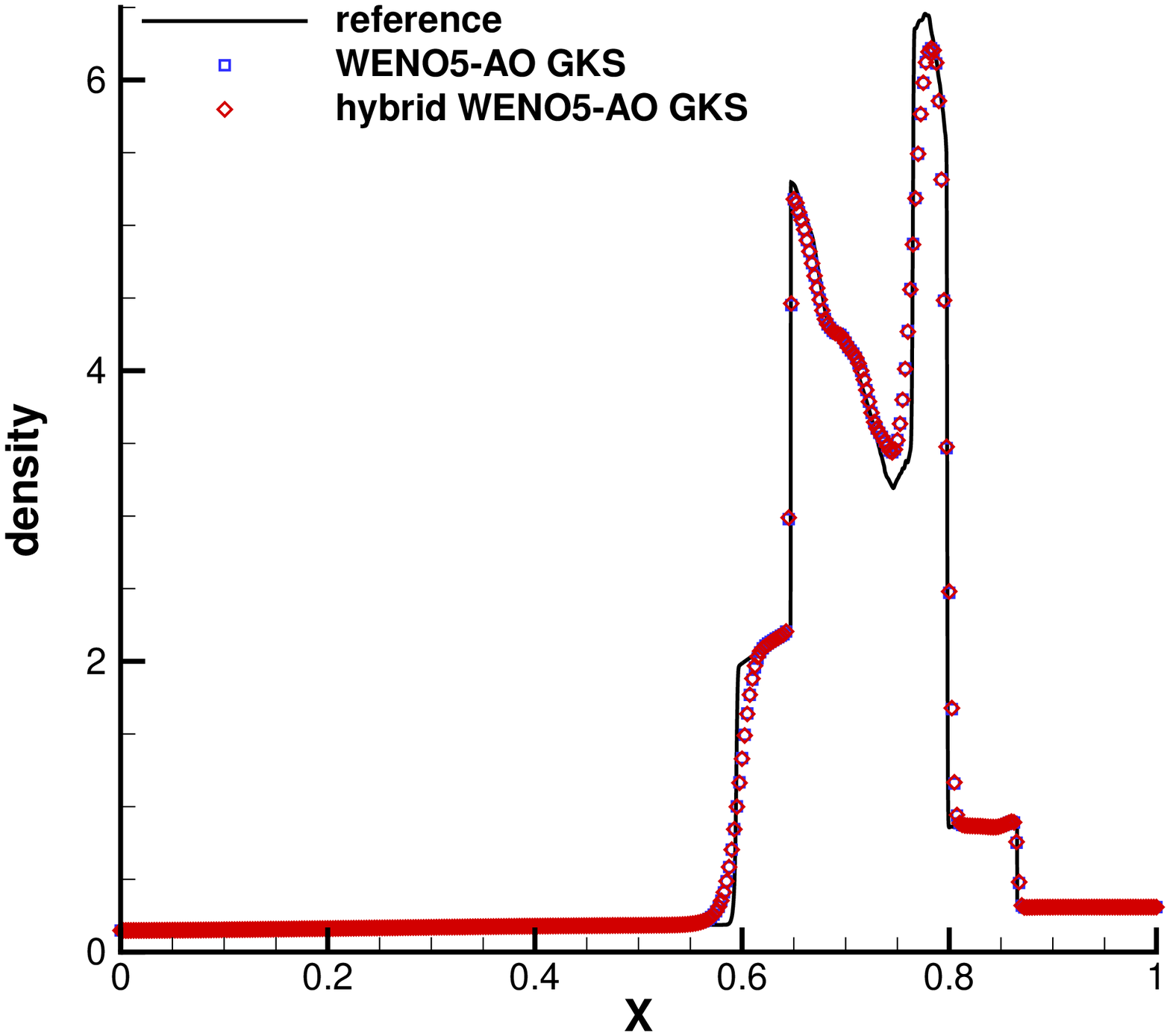}}
	\subfigure{\includegraphics[width=0.48\textwidth]{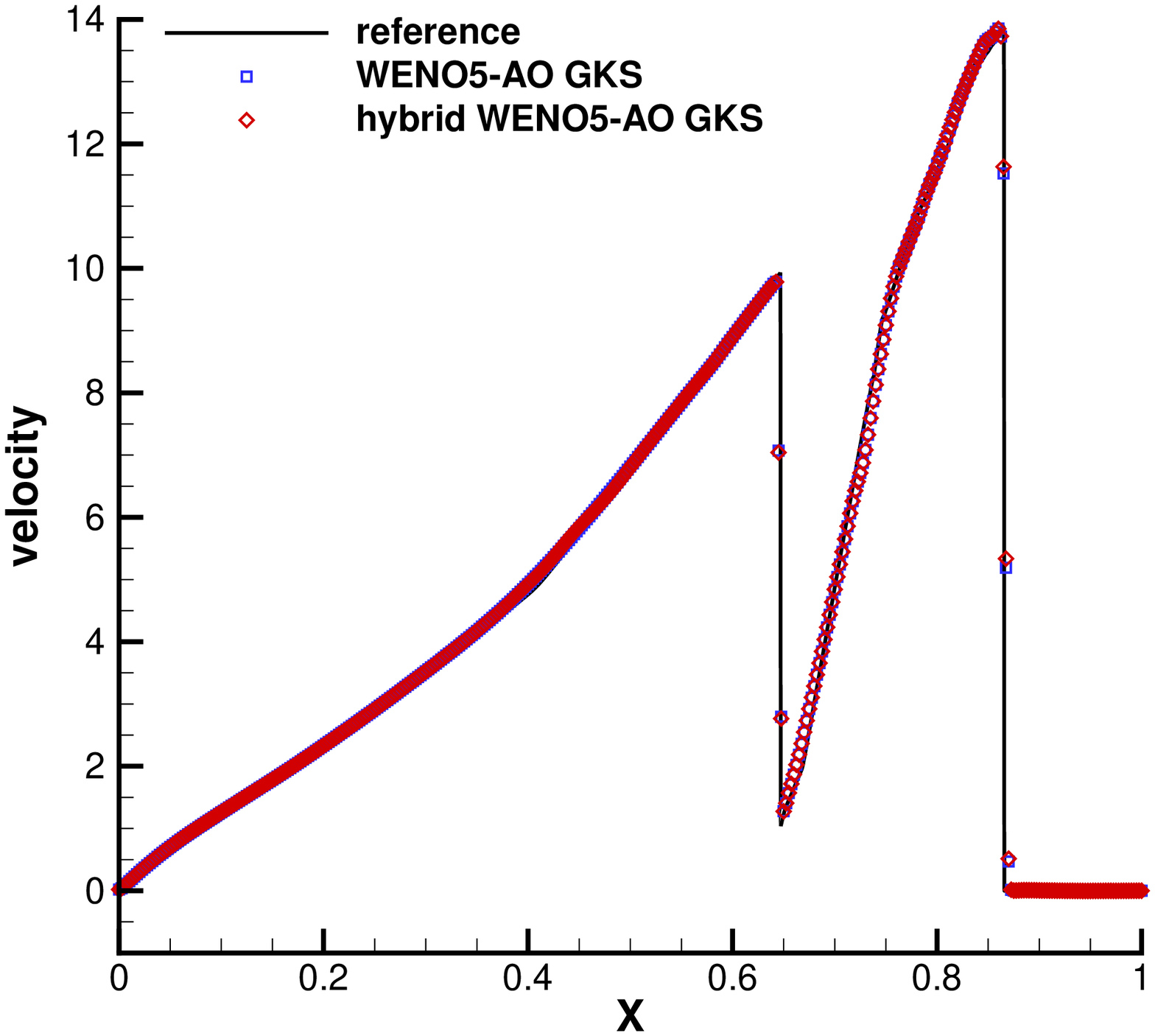}}
	\subfigure{\includegraphics[width=0.48\textwidth]{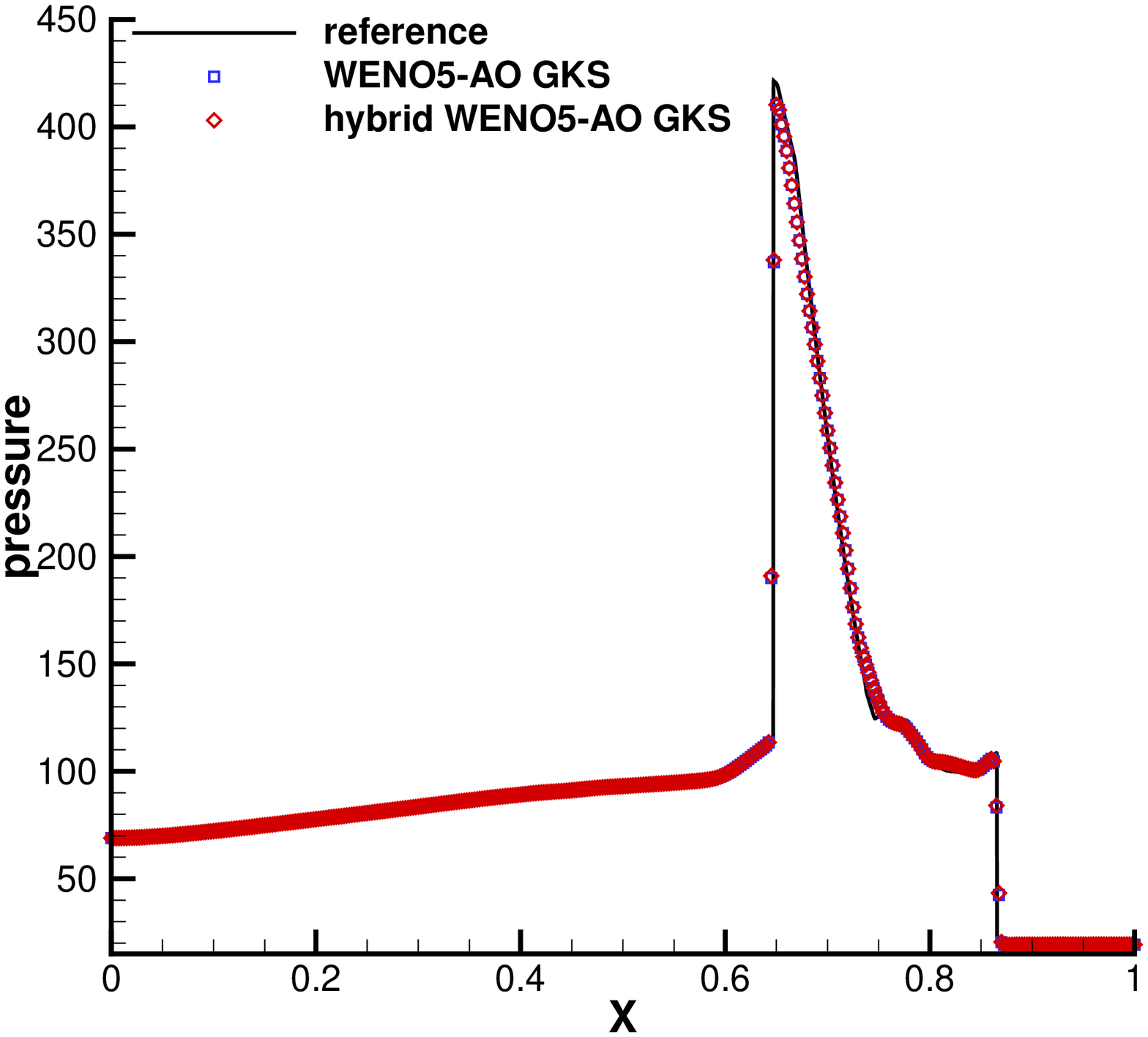}}
	\caption{Blast wave problem: the density, velocity and pressure distributions for WENO-AO reconstruction and hybrid WENO-AO reconstruction with 400 cells.  $CFL=0.5.$ $T=0.038.$ }
	\label{figblast}
\end{figure}

The computation is conducted with $CFL=0.5$, employing 400 equally spaced grid points while applying reflection boundary conditions at both ends. Fig. \ref{figblast} displays the computed profiles of density, velocity and pressure at t=0.038. The results of HGKS with hybrid WENO-AO reconstruction and HGKS with WENO-AO reconstruction show identical wave profiles and correspond well with the reference solutions. The HGKS with hybrid WENO-AO reconstruction exhibits a 75.66\% reduction in CPU time for the reconstruction procedure. This test case contains strong discontinuities requiring high-order schemes to be robust. These results indicate that the hybrid WENO5-AO GKS is as robust as the WENO5-AO GKS.\\

\subsection{Accuracy test in 2-D}
\label{sec3.3}
Similar to 1-D case, we choose the advection of density perturbation for accuracy test. The collision time $\tau =0$ is set for this inviscid flow. The initial conditions are 
$$\rho \left( x,y \right)=1+0.2sin\left( \pi \left( x+y \right) \right),U\left( x,y \right)=\left( 1,1 \right),p\left( x,y \right)=1,\left( x,y \right)\in \left[ 0,2 \right]\times \left[ 0,2 \right].$$
The compute domain is $\left[ 0,2 \right]\times \left[ 0,2 \right]$. $N\times N$ uniform mesh is used and the boundary conditions in both directions are periodic. The analytic solution of the 2-D advection of density perturbation is
$$\rho \left( x,y,t \right)=1+0.2sin\left( \pi \left( x+y-t \right) \right),U\left( x,y,t \right)=1,p\left( x,y,t \right)=1,\left( x,y \right)\in \left[ 0,2 \right]\times \left[ 0,2 \right].$$
The CFL number of the time steps are 0.5. HGKS with both WENO-AO reconstruction and hybrid WENO-AO reconstruction are tested and presented in the Table \ref{2Daccuracy01} and Table \ref{2Daccuracy02} respectively. The HGKS with hybrid WENO-AO reconstruction can achieve expected accuracy as the HGKS with WENO reconstruction.
\begin{table}
\begin{tabular}{c|cc|cc|cc}
	\hline
	mesh length & ${{L}^{1}}$error & Order & ${{L}^{2}}$error & Order & ${{L}^{\infty }}$error & Order\\
	\hline
	1/5 & 1.3573113e-03 &  & 1.4897997e-03 &  & 2.1046216e-03 & \\ 
	1/10 & 4.1633737e-05 & 5.03 & 4.6328748e-05 & 5.01 & 6.7573520e-05 & 4.96 \\
	1/20 & 1.3564842e-06 & 4.94 & 1.5063088e-06 & 4.94 & 2.1917062e-06 & 4.95\\
	1/40 & 4.8651945e-08 & 4.80 & 5.4154814e-08 & 4.80 & 7.7820892e-08 & 4.82\\
	1/80 & 1.7467843e-09 & 4.80 & 1.9404552e-09 & 4.80 & 2.8182315e-09 & 4.79\\
	\hline
\end{tabular}
\centering 
	\caption{\label{2Daccuracy01}Accuracy test in 2-D for the advection of density perturbation by the WENO-AO reconstruction. $CFL=0.5$}
\end{table}
\begin{table}
\begin{tabular}{c|cc|cc|cc}
	\hline
	mesh length & ${{L}^{1}}$error & Order & ${{L}^{2}}$error & Order & ${{L}^{\infty }}$error & Order\\
	\hline
	1/5 & 1.8574494e-03 &  & 2.0349011e-03 &  & 2.9200462e-03 & \\ 
	1/10 & 5.8628718e-05 & 4.99 & 6.5166053e-05 & 4.96 & 9.4883564e-05 & 4.94 \\
	1/20 & 1.8571539e-06 & 4.98 & 2.0587183e-06 & 4.98 & 3.0099476e-06 & 4.98\\
	1/40 & 6.0991328e-08 & 4.93 & 6.7640692e-08 & 4.93 & 9.8766773e-08 & 4.93\\
	1/80 & 2.0681312e-09 & 4.88 & 2.2970915e-09 & 4.88 & 3.6052416e-09 & 4.78\\
	\hline
\end{tabular}
\centering 
	\caption{\label{2Daccuracy02}Accuracy test in 2-D for the advection of density perturbation by the hybrid WENO-AO reconstruction. $CFL=0.5$}
\end{table}
For the following 2-D Numerical examples, the CPU time of the HGKS with different reconstruction procedure is presented in Table \ref{2DCPU}, in which only the CPU time of the spatial reconstructor is counted. Similarly, in the double Mach reflection problem with the same initial conditions, the CPU time of the remaining programs for the two schemes is 45235.455s and 45478.215s, respectively. So we still show the effect of the hybrid WENO-AO method on HGKS by comparing the CPU time of spatial reconstruction.
\begin{table}
\begin{tabular}{p{4cm}|c|p{4cm}|p{3cm}|c}
	\hline
	Numerical example & Grid points & CPU time of hybrid WENO-AO (s) & CPU time of WENO-AO (s) & Ratio\\
	\hline
	Riemann problem (configuration 1) & $500\times 500$ & 706.471 & 6100.513 & 11.58\% \\ 
	Riemann problem (configuration 6) & $500\times 500$ & 5546.881 & 9628.309 &57.61\%  \\
	Double Mach reflection problem & $960\times 240$ & 5334.322 & 13596.93 & 39.23\% \\
	Viscous shock tubes problem & $500\times 250$ & 12148.766 & 46394.377 & 26.91\% \\
	\hline
\end{tabular}
\centering 
	\caption{\label{2DCPU}Computing time of the hybrid WENO-AO reconstruction procedure and the WENO-AO reconstruction procedure in 2-D test cases.}
\end{table}
\subsection{Two-dimensional Riemann problems}
\label{sec3.4}
\emph{(a) Configuration 1}\par
\begin{figure}[htbp]
	\centering
	\subfigure{\includegraphics[width=0.4\textwidth]{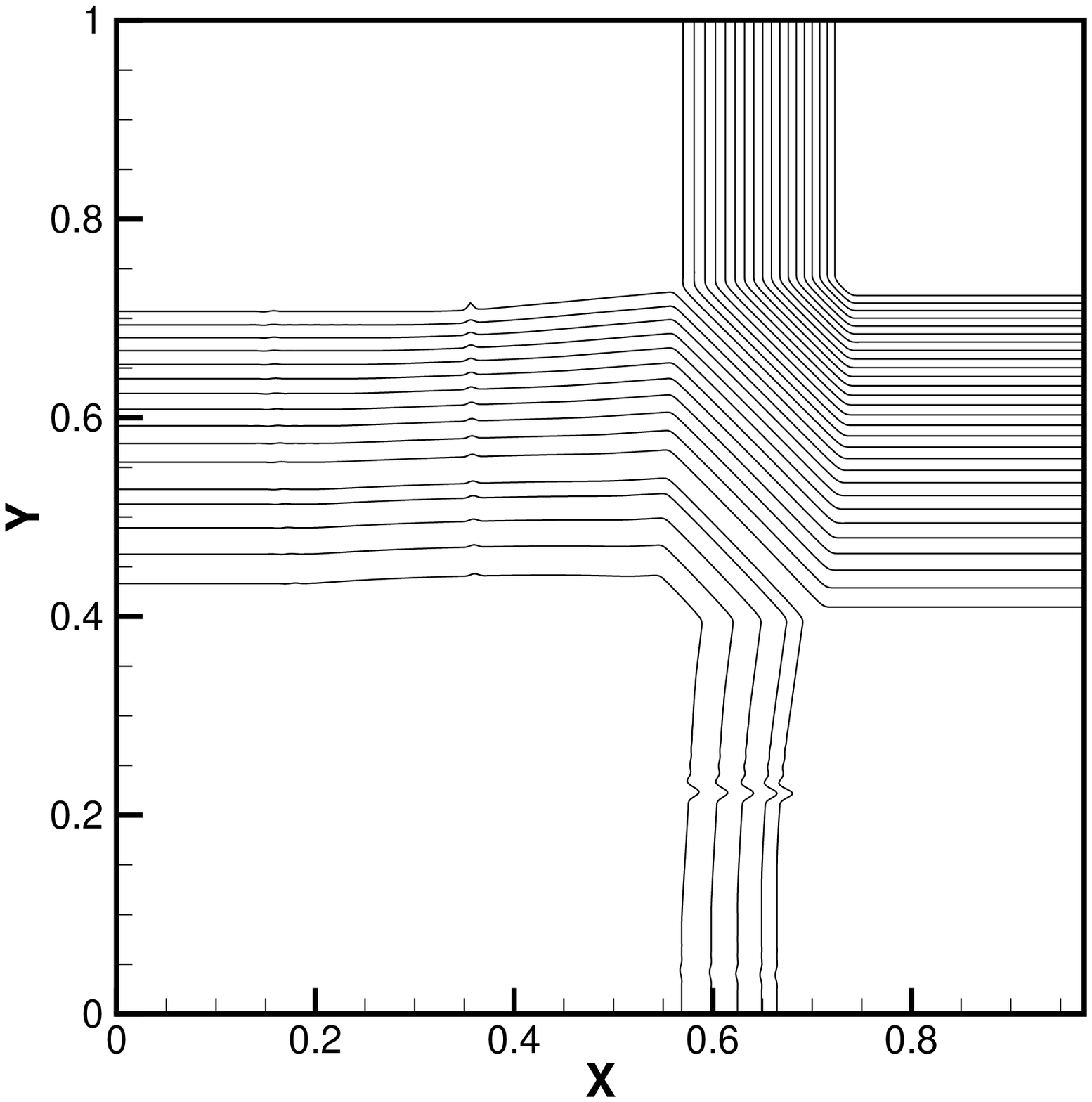}}
	\subfigure{\includegraphics[width=0.4\textwidth]{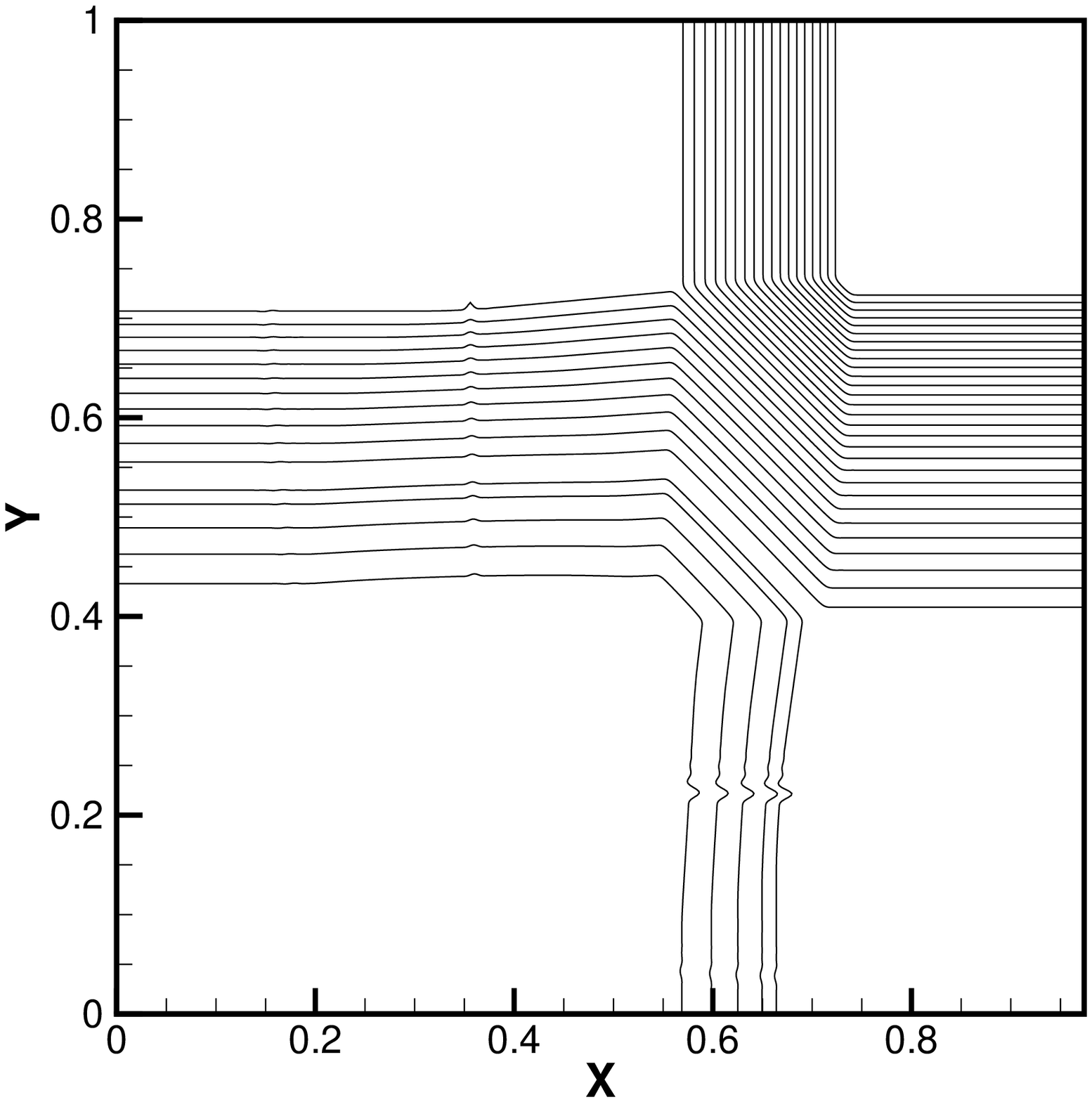}}
	\caption{Two-dimensional Riemann problems: the density distributions for Configuration 1. Left: the HGKS with WENO-AO scheme. Right: the HGKS with hybrid WENO-AO scheme. $CFL=0.5.$ $T=0.2.$ Mesh:$500\times 500.$}
	\label{figliman01}
\end{figure}

The initial conditions of the Configuration 1 are four 1-D rarefaction waves and given as~\cite{SOTR}
$$\left( \rho ,U,V,p \right)=\left\{ \begin{aligned}
  & \left( 0.1072,-0.7259,-1.4045,0.0439 \right),x<0.5,y<0.5, \\ 
 & \left( 0.2579,0,-1.4045,0.15 \right),x\ge 0.5,y<0.5, \\ 
 & \left( 1,0,0,1 \right),x\ge 0.5,y\ge 0.5, \\ 
 & \left( 0.5197,-0.7259,0,0.4 \right),x<0.5,y\ge 0.5. \\ 
\end{aligned} \right.$$
The results of the Configuration 1 at $t=0.2$ for the HGKS with WENO-AO method and hybrid WENO-AO method are presented in Fig. \ref{figliman01}. In comparison to the HGKS with WENO-AO method ~\cite{PEFH}, the original HGKS scheme yields tangential reconstructions without negative temperature due to the separated smooth tangential reconstructions for the equilibrium state ${{g}^{c}}$. Although the HGKS with hybrid WENO-AO method also introduces linear reconstruction in tangential reconstructions. The HGKS with hybrid WENO-AO method has also no negative temperature in this case. So it has the same stability as the HGKS with the WENO-AO reconstruction, but is more efficient.\\
\emph{(b) Configuration 6}\par
\begin{figure}[htbp]
	\centering
	\subfigure{\includegraphics[width=0.48\textwidth]{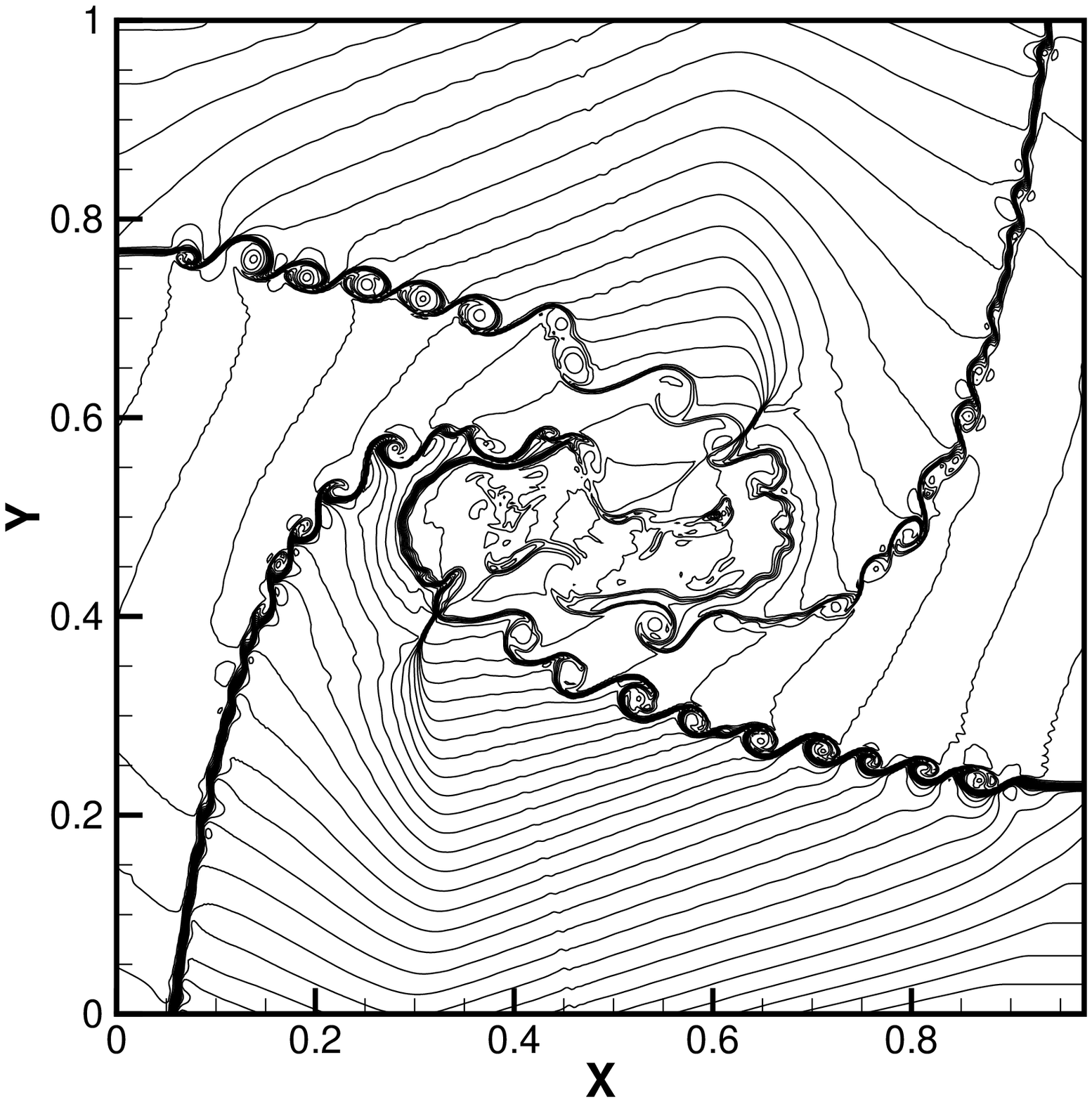}}
	\subfigure{\includegraphics[width=0.48\textwidth]{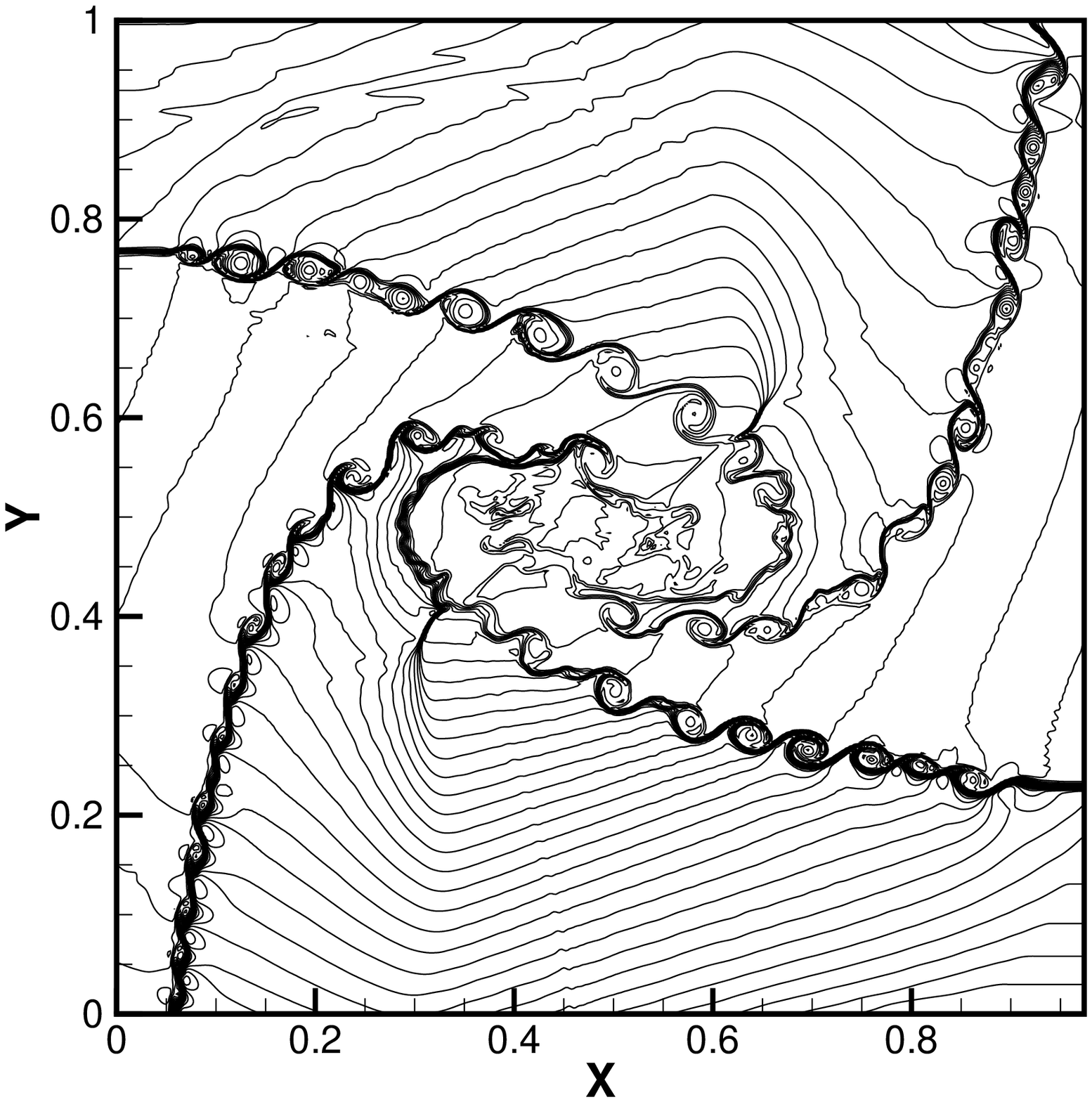}}
	\caption{Two-dimensional Riemann problems: the density distributions for Configuration 6. Left: the HGKS with WENO-AO scheme. Right: the HGKS with hybrid WENO-AO scheme. $CFL=0.95.$ $T=0.6.$ Mesh:$500\times 500.$}
	\label{figliman02}
\end{figure}

For compressible flow, the shear layer flow is one of the most distinguishable flow patterns. For the case of Configuration 6 ~\cite{SOTR}, the initial conditions with four planar contact discontinuities are given as  
 $$\left( \rho ,U,V,p \right)=\left\{ \begin{aligned}
  & \left( 1,-0.75,0.5,1 \right),x<0.5,y<0.5, \\ 
 & \left( 3,-0.75,-0.5,1 \right),x\ge 0.5,y<0.5, \\ 
 & \left( 1,-0.75,-0.5,1 \right),x\ge 0.5,y\ge 0.5, \\ 
 & \left( 2,0.75,0.5,1 \right),x<0.5,y\ge 0.5. \\ 
\end{aligned} \right.$$
These discontinuities trigger the K-H instabilities due to the numerical viscosities. It is commonly believed that the less numerical dissipation corresponds to larger amplitude shear instabilities ~\cite{EORF}. It can be observed in Fig. \ref{figliman02} that the HGKS with hybrid WENO-AO method predicts less numerical dissipation and more details of vortices than the HGKS with WENO-AO method. This is mainly due to the fact that in a few cases, direct upwind linear reconstruction has better performance than WENO-AO reconstruction. Also as shown in Table \ref{2DCPU}, the HGKS with hybrid WENO-AO method can save about 50\% CPU of the reconstruction procedure.\par
The HGKS with hybrid WENO-AO reconstruction can save 46.03\% more CPU time reduction in Configuration 1 than in Configuration 6. Because the flow of the Configuration 6 is more complex than Configuration 1. The proportion of WENO-AO reconstruction increases in the Configuration 6 during the reconstruction of the conservative variables and its derivatives in their normal and tangential directions. What is the same is that their initial conditions are all four parts of different flow region, but the computational cost reduction is somewhat different.\\

\subsection{Double Mach reflection problem}
\label{sec3.5}
\begin{figure}
	\centering
	\subfigure{\includegraphics[width=0.74\textwidth]{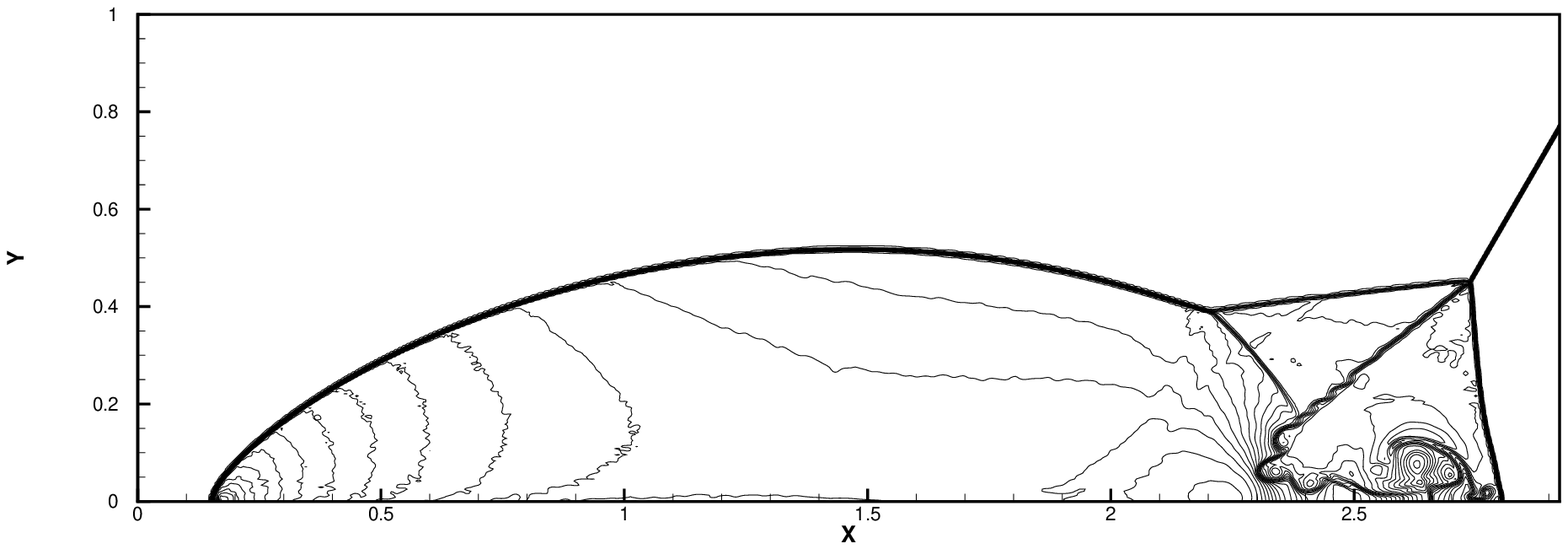}}
	\subfigure{\includegraphics[width=0.25\textwidth]{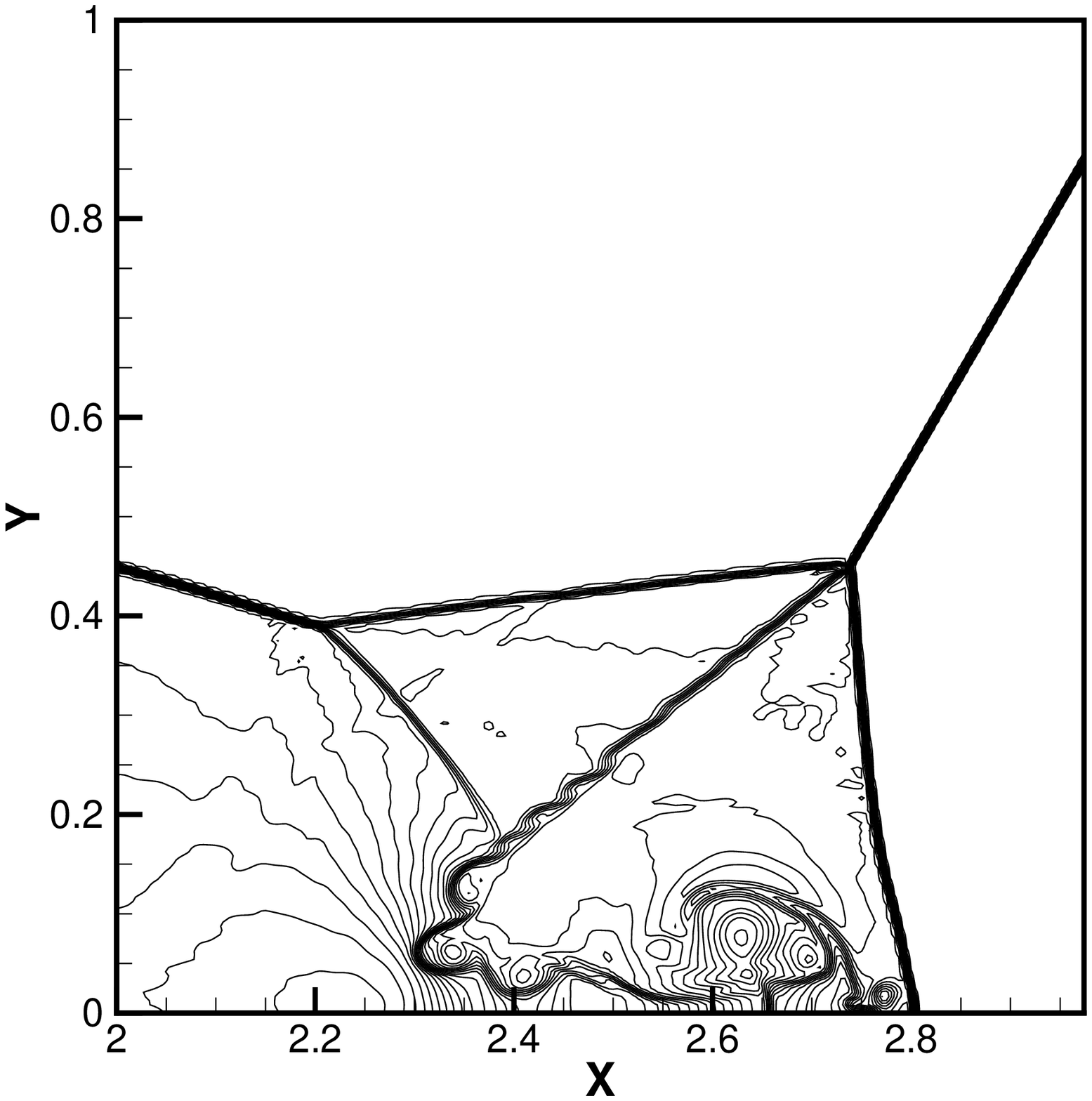}}
	\caption{The density distributions for Double Mach: the HGKS with hybrid WENO-AO scheme. Mesh: $960\times 240.$ $CFL=0.8.$ $T=0.2.$ ${{c}_{1}}=0.$${{c}_{2}}=1.$ }
	\label{figdoublehy}
\end{figure}

\begin{figure}
	\centering
	\subfigure{\includegraphics[width=0.74\textwidth]{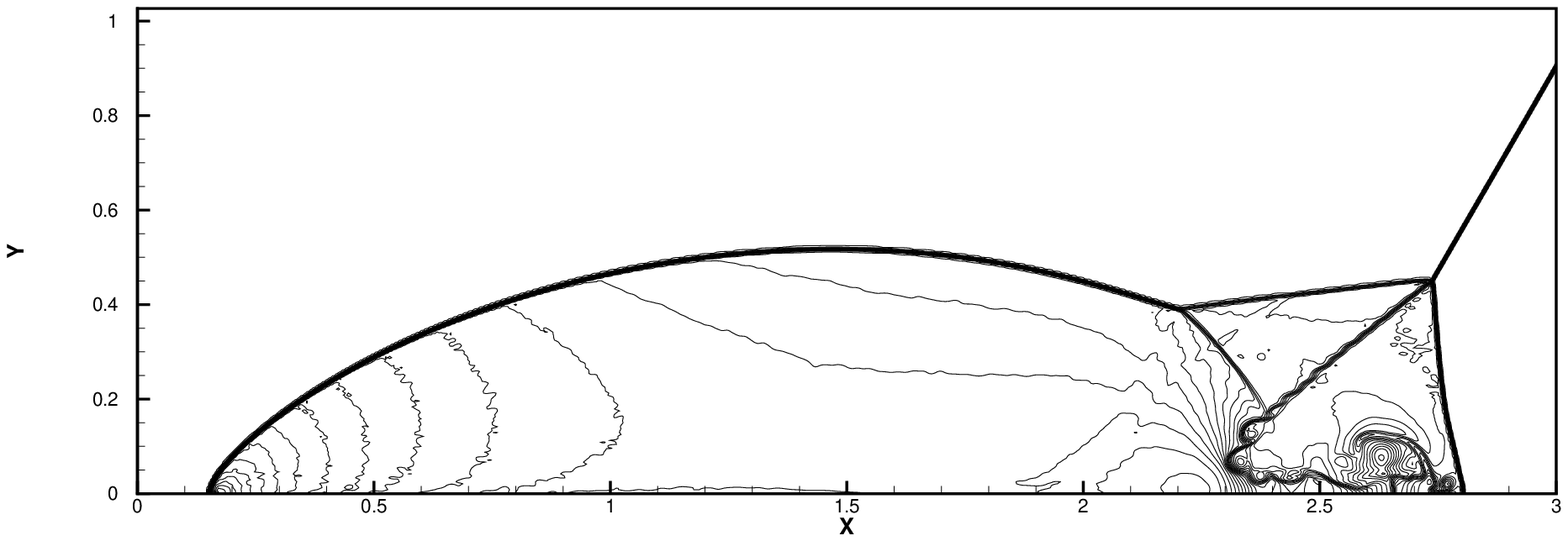}}
	\subfigure{\includegraphics[width=0.25\textwidth]{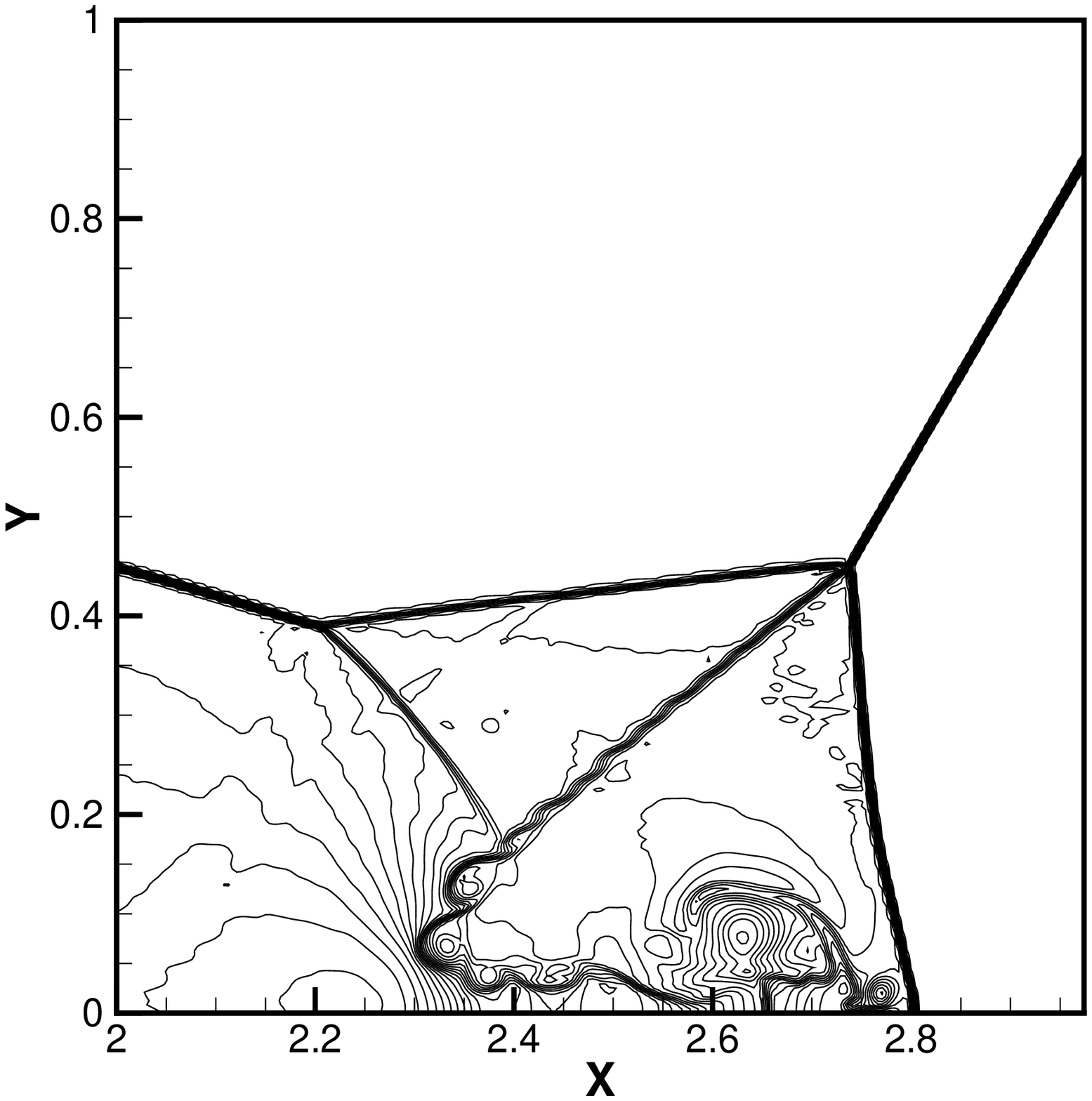}}
	\caption{The density distributions for Double Mach: the HGKS with WENO-AO scheme. Mesh: $960\times 240.$ $CFL=0.8.$ $T=0.2.$ ${{c}_{1}}=0.$${{c}_{2}}=1.$ }. 
	\label{figdoublenohy}
\end{figure}

The double Mach reflection problem is a inviscid test case designed by Woodward and Colella~\cite{TNSO}. The computational domain is $\left[ 0,\ 4 \right]\times \left[ 0,\ 1 \right]$ with a slip boundary condition applied on the bottom of the domain starting from $x=1/6$. The post -shock condition is set for the rest of bottom boundary. At the top boundary, the flow variables are set to describe the exact motion of the Mach 10 shock. The initial pre-shock and post-shock conditions are
$$\begin{aligned}
  & \left( \rho ,U,V,p \right)=\left( 8,4.125\sqrt{3},-4.125,116.5 \right), \\ 
 & \left( \rho ,U,V,p \right)=\left( 1.4,0,0,1 \right). \\ 
\end{aligned}$$
Initially, a right-moving Mach 10 shock with a ${{60}^{\circ }}$ angle against the x-axis is positioned at $\left( x,y \right)=\left( 1/6,0 \right)$.
	The density distributions and local enlargements at $t=0.2$ for the new HGKS with hybrid WENO-AO scheme and the HGKS with WENO-AO scheme are plotted in Fig.\ref{figdoublehy} and Fig.\ref{figdoublenohy}, respectively. The HGKS with hybrid WENO-AO method and the HGKS with WENO-AO method are robust and well validated with increasing the CFL number up to 0.8. But the classic WENO5 GKS scheme is With validated with $CFL=0.5$.  The HGKS with hybrid WENO-AO method can save 60\% CPU time of the reconstruction procedure. Besides the new HGKS with hybrid WENO-AO scheme also resolves the flow structure under the triple Mach stem clearly as the WENO5-AO GKS. 

\subsection{Viscous shock tubes problem}
\label{sec3.6}
This viscous shock tube problem~\cite{CHCS2,AEAM,NSOT2} was investigated to valid the capability of the present two HGKS schemes for low Reynolds number viscous flow with strong shocks. In a two-dimensional unit box $\left[ 0,\ 1 \right]\times \left[ 0,\ 1 \right]$, a membrane located at $x=0.5$ separates two different states of the gas and the dimensionless initial states are 
$$\left( \rho ,U,p \right)=\left\{ \begin{aligned}
  & \left( 120,0,120/\gamma  \right),0<x<0.5, \\ 
 & \left( 1.2,0,1.2/\gamma  \right),0.5<x<1, \\ 
\end{aligned} \right.$$
where $\gamma =1.4$, Prandtl number $\Pr =1.0$ and Reynolds number $\operatorname{Re}=1/\mu =200$. The computational domain is $\left[ 0,\ 1 \right]\times \left[ 0,\ 0.5 \right]$ with a symmetric boundary condition imposed on the top $x\in \left[ 0,\ 1 \right]$, $y=0.5$ and non-slip adiabatic conditions applied on the other three solid wall boundaries.\par 
At $t=0$, the membrane is removed and a wave interaction occurs. A shock wave with Mach number $Ma=2.37$ moves to the right side. Then the shock wave followed by a contact discontinuity reflects at the right wall. After the reflection, the shock interacts with the contact discontinuity. During their propagation, both of them interact with the horizontal wall and create a thin boundary layer. The solution will develop complex two-dimensional shock/shear/boundary-layer interactions and the dramatic changes for velocities above the bottom wall introduce strong shear stress.\par
\begin{figure}
	\centering
	\includegraphics[width=1.0\textwidth]{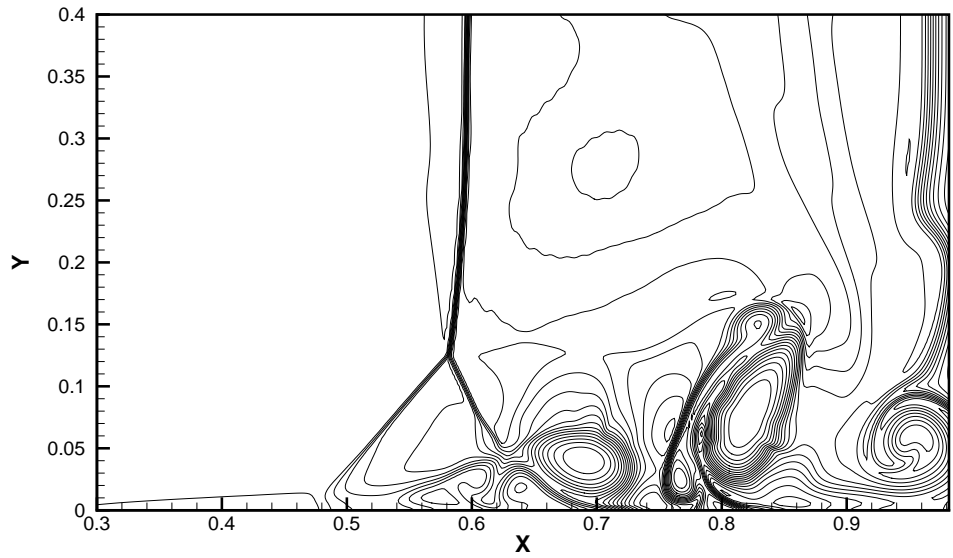}
	\caption{\label{figviscousnohy}The density contours at $t=1$ for $Re=200$ viscous shock tube: the HGKS with WENO-AO scheme. Mesh: $500\times 250.$ $CFL=0.3.$}
\end{figure}

\begin{figure}
	\centering
	\includegraphics[width=1.0\textwidth]{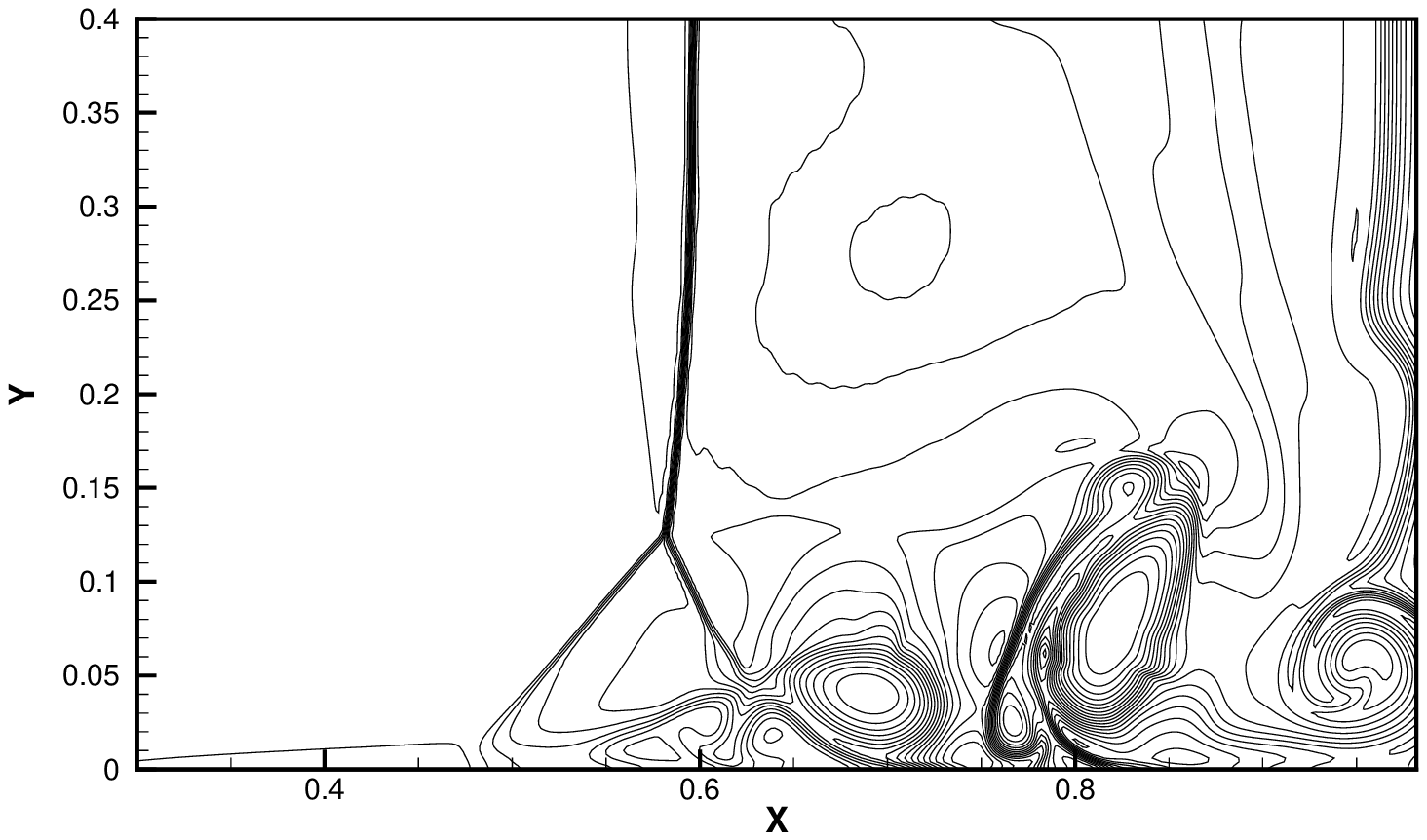}
	\caption{\label{figviscoushy}The density contours at $t=1$ for $Re=200$ viscous shock tube: the HGKS with hybrid WENO-AO scheme. Mesh: $500\times 250.$ $CFL=0.3.$}
\end{figure}

\begin{figure}
	\centering
	\includegraphics[width=1.0\textwidth]{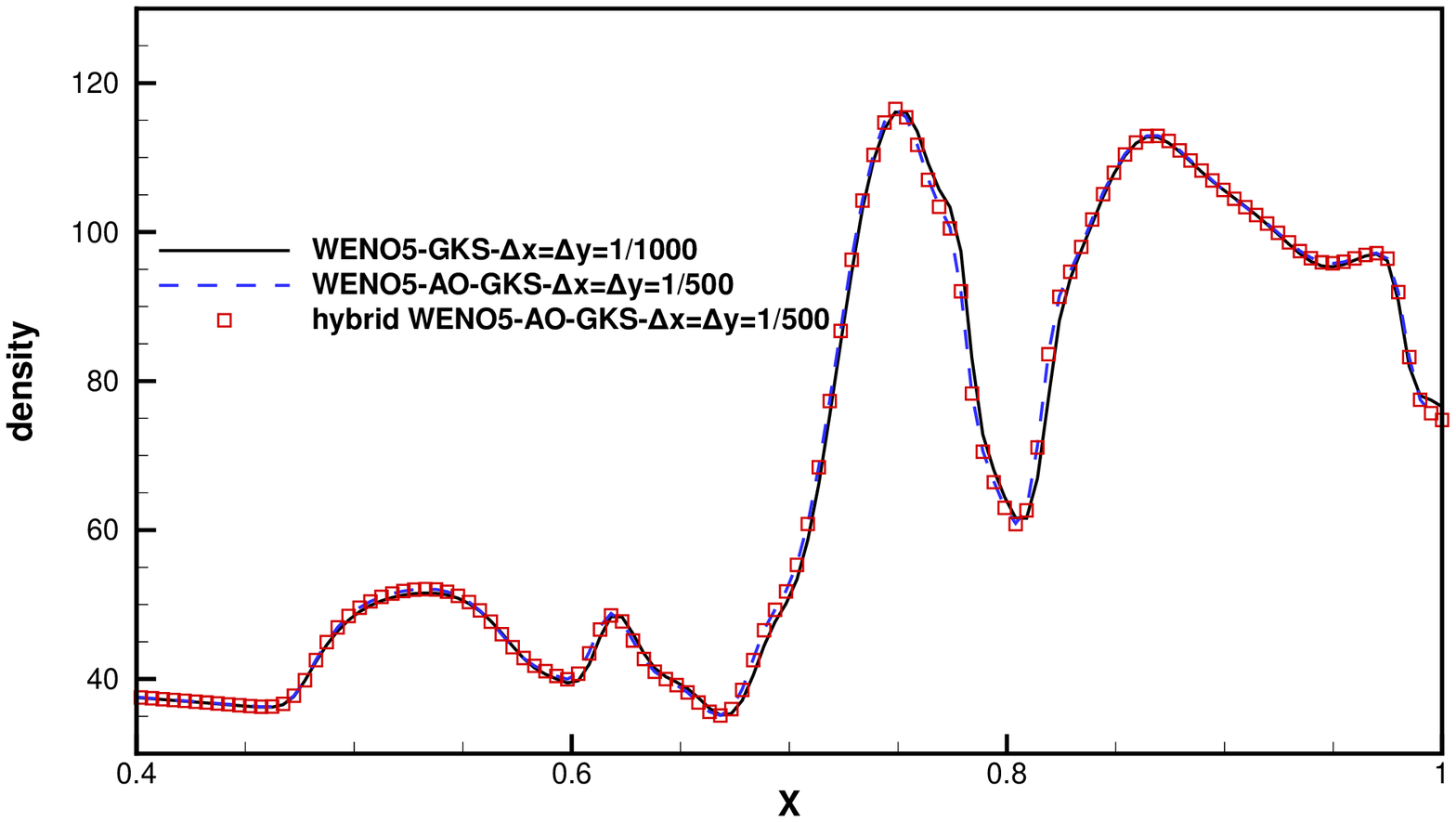}
	\caption{\label{figviscousdensity}The density profiles along the bottom wall at $t=1$ for $Re=200$ viscous shock tube.}
\end{figure}

 
The density distributions of viscous shock tube at $t=1.0$ for the HGKS with WENO-AO method is plotted in Fig. \ref{figviscousnohy}. And the density distributions of viscous shock tube at $t=1.0$ for the HGKS with hybrid WENO-AO method is plotted in Fig. \ref{figviscoushy}. The density profiles along the bottom wall are shown in Fig. \ref{figviscousdensity}. The classic HGKS replaces WENO-Z type weights with WENO-JS weights to pass this test case. However, both the HGKS with WENO-AO in ~\cite{PEFH} and the new HGKS with hybrid WENO-AO method can pass and the obtain the following results. Compared with the HGKS with WENO-AO method, the new HGKS with hybrid WENO-AO method can save about 75\% of CPU time for reconstruction procedure. In Fig. \ref{figviscousdensity}, the density profiles of HGKS with WENO-AO and hybrid WENO-AO are almost same and agree well with the density profiles of classic WENO5-GKS with $\Delta x=\Delta y=1/1000$.

\section{Conclusion}
\label{sec4}
The two-stage fourth-order gas-kinetic scheme (HGKS) based on the hybrid WENO-AO method retains the high accuracy and robustness of the original WENO5-AO GKS. The troubled cell indicator in the hybrid WENO-AO method provides a reliable way to determine if all extreme points of the fourth-degree polynomial reconstructed in the cell or interface are within the large stencils. Moreover, the slopes of the flow variables are derived from the reconstructed nonlinear or linear polynomials. HGKS with the hybrid WENO-AO method reduces spurious oscillations at weak discontinuities and exhibits good shear instability resolution, as verified by various inviscid and viscid numerical tests. Importantly, compared to the HGKS with the WENO-AO method, the computation time of the HGKS spatial reconstruction procedure with the hybrid WENO-AO method is reduced by 80\% in 1D and 60\% in 2D, while maintaining similar robustness and accuracy. Futhermore, the computational cost of the three-dimensional spatial reconstruction procedure of HGKS is substantially higher than that of two-dimensional spatial reconstruction. Therefore, future work will continue to verify whether the HGKS with hybrid WENO-AO method can ensure high accuracy and robustness while enhancing computational efficiency. Our goal is to employ HGKS with hybrid WENO-AO method in numerical simulations of three-dimensional complex flow.

%%%% Acknowledgments %%%%%%%%
\section*{Acknowledgments}
The authors would like to thank Yining Yang, Hao Jin for helpful discussion.
This work is sponsored by the National Natural Science Foundation of China [grant numbers 11902264, 12072283, 12172301] and the 111 Project of China (B17037).

%%%% Bibliography  %%%%%%%%%%

\end{document}